\documentclass[12pt]{article}

\usepackage{amsfonts,amssymb,amsmath,latexsym,epsfig}

\usepackage{cite, euscript}
\newcommand{\pd}{\partial}

\DeclareMathOperator{\sign}{sign}

\newcommand{\Dd}{\EuScript{D}^2}
\newcommand{\Dh}{\Dd_\mathrm{H}}

\newcommand{\DD}[1]{\Dd_{\mathrm{H}_{#1}}}

\newcommand{\xio}{\ensuremath{\xi_{\text{osc}}^2}}
\newcommand{\xis}{\ensuremath{\xi_{\text{shape}}^2}}

\newcommand{\xip}{\ensuremath{\xi_{\text{phys}}^2}}

\title{
\vskip -50pt
{\begin{normalsize}
\mbox{} \hfill DAMTP-2007-67 \\
\vskip 50pt
\end{normalsize}}
Dynamics in Nonlocal Cosmological Models Derived from String Field Theory
}

\author{Liudmila Joukovskaya\thanks{E-mail: \texttt{l.joukovskaya@damtp.cam.ac.uk}}\\
DAMTP, Centre for Mathematical Sciences, University of Cambridge,\\
Wilberforce Road, Cambridge CB3 0WA, UK}

\date{}

\begin{document}

\maketitle

\begin{abstract}
A general class of nonlocal cosmological models is considered. A
new method for solving nonlocal Friedmann equations is proposed,
and solutions of the Friedmann equations with nonlocal
operator are presented. The cosmological properties of these
solutions are discussed. Especially indicated is $p$-adic
cosmological model in which we have obtained nonsingular
bouncing solution and string field theory tachyon model in which
we have obtained full solution of nonlocal Friedmann equations
with $w=-1$ at large times. The possibility of obtaining
realistic value of cosmological constant from nonlocal
cosmological models is also discussed.
\end{abstract}

\section{Introduction}

In recent works there appears interest in nonlocal cosmological
models derived from String Field Theory in connection with
problem of describing cosmological inflation or accelerating
expansion of the Universe.

Modern cosmological data indicates that expansion of the
Universe is accelerating. It may be owing to a component of the
Universe with negative pressure, Dark Energy. Recent results of
WMAP \cite{cosmo-obser} together with Ia supernovae data give
us the following range for the dark energy state parameter
$w=-0.97^{+0.07}_{-0.09}$. In the assumption that the value of
the state parameter changes in the range $0>w>-1$ there exist
different theoretical models for the dark energy (see reviews \cite{DE1,DE2}
and references therein)
--
quintessence models \cite{quintessence}, K-essence\cite{k-essence},
 quintom models \cite{quintom}, models constructed on the DBI action (see reviews
\cite{GG} and references therein, \cite{tachyonmodels}) and dilatonic models
\cite{dilatonicmodels}. The case
$w=-1$ corresponds to models with cosmological constant, which
are also possible. We also can find in the literature models
related $w<-1$ with modified gravity \cite{modifiedgravity}.

A nontrivial possibility is the case $w<-1$, that means the violation of the
null energy condition (NEC). One of the possibilities is to consider the
phantom Universe \cite{IA_marion, phantom}.
There were proposed several
coupled scalar-gravity models in which the null energy condition is violated,
but as an effect most such models are unstable\cite{scalargravitymodels}.
Recently it was proposed another possibility to consider dark energy description in
the context of $D$-brane decay in cubic superstring field theory
\cite{IA_marion}.  It was shown that $D$-brane decay can be
interpreted at least at late times as a phantom model. Note
that unlike phenomenological phantom models here phantom is an
effective theory. Since string field theory by itself is a
consistent  quantizable theory this approach does not suffer from
instabilities which are inevitable  for phenomenological
phantom models \cite{IA_marion}. Such a nonlocal cosmological
models were also considered in \cite{AJ-JHEP,AJV-JHEP,Calcagni,Lump,CMN}.

At the same time there have been a number of attempts to
realize description of the early Universe via nonlocal
cosmological models \cite{BarnabyBiswasCline,
BarnabyCline,Lidsey,CMN}. One example is $p$-adic inflation
model \cite{BarnabyBiswasCline} which is represented by
nonlocal $p$-adic string theory coupled to gravity. For this
model, a  rolling inflationary solution was constructed and the
interesting features were discussed and compared with cosmic
microwave background (CMB) observations. The possibility of
obtaining large nongaussian signatures in the CMB
has also been considered in a general class of single field
nonlocal hill-top inflation models \cite{BarnabyCline}. Another
example is investigation of the inflation near a maximum of the
nonlocal potential when non-local derivative operators are
included in the inflaton Lagrangian. It was found that
higher-order derivative operators in the inflaton Lagrangian
can support a prolonged phase of slow-roll inflation near a
maximum of the potential \cite{Lidsey}.

It may happen that  both early stage and contemporarily
Universe can be explained with a single nonlocal model derived
from string field theory. To be able to tell
more about the early stage we need to consider perturbations in
nonlocal models. It will be very interesting to compare
obtained results in particular with those in \cite{Mulryne,
Wesley, LFTS}. To be able to do it the new technique for
solving nonlinear Friedmann equation is required which will
hopefully be provided in the present work where for the first
time construction of solution of the full nonlocal Friedmann
equations is presented. That is why in the present work we
discuss numerical solution of Friedmann equations and their
features in more detail.

One of the popular examples of nonlocal models is the $D$-brane
decay in the gravitational background in the framework of string
field theory which is described by the following tachyon action
\begin{equation}
\label{tach-ac}
S=\int d^4x\sqrt{-g}\left(\frac{m_p^2}{2}R+
\frac{\xi^2}{2} \phi\square_g\phi+\frac{1}{2}\phi^2-
\frac{1}{4}\Phi^4-T-\Lambda^\prime \right),
\end{equation}
where $g$ is the metric, $m_p$ is reduced Planck mass, $\phi$
is tachyon field, $\Phi=e^{k \Box}\phi$, $k=\frac{1}{8}$,
$\Box_g$ is usual Beltrami-Laplace operator, $T$ is tension of
the $D3$ brane, $\Lambda^{\prime}$ is the effective cosmological constant.

Note that covariant string field theory as of today is
constructed only in the flat background. It was built only for
some special types of non-flat backgrounds, in particular, in
the anti-de sitter background \cite{Metsaev1} and in the plane
wave background \cite{Metsaev2}. So the action (\ref{tach-ac})
at the special choice of $V(\Phi)$ is a direct generalization
for the approximated tachyon action to the case of arbitrary
metric.

The action (\ref{tach-ac}) in the Friedmann-Robertson-Walker (FRW) background
$$
ds^2=-dt^2+a^2(t)(dx_1^2+dx_2^2+dx_3^2)
$$
leads to the following Friedmann equations
\begin{equation}
3H^2=\frac{1}{m_p^2}~{\cal E},
\end{equation}
\begin{equation}
3H^2+2\dot H={}-\frac{1}{m_p^2}~{\cal P},
\end{equation}
which have the same form as the usual equations except for extra term in the
expression for the energy $\mathcal{E}$ and pressure
$\mathcal{P}$ which appear from the nonlocal interaction and which
contain the Hubble function. This term makes these Friedmann
equations more complicated even for numerical consideration.
More precisely,
\begin{eqnarray}
\mathcal{E}&=
\underbrace{\mathcal{E}_k+\mathcal{E}_p+T+\Lambda^\prime}_{
  \text{\textit{usual local energy expression}}}
+~~\mathcal{E}_{nl1}~+~\mathcal{E}_{nl2},\\
\mathcal{P}&=
\underbrace{\mathcal{E}_k-\mathcal{E}_p-T-\Lambda^\prime}_{
  \text{\textit{usual local pressure expression}}}
-~~\mathcal{E}_{nl1}~+~\mathcal{E}_{nl2},
\end{eqnarray}
where
$$
{\cal E}_{nl1}=k\int_{0}^{1} d \rho \left(e^{k\rho \Box_g}\Phi^3 \right)
 \left(-\Box_g e^{-k\rho\Box_g}\Phi\right),
~~{\cal
E}_{nl2}=-k\int_{0}^{1}  d \rho\left(\partial e^{k \rho\Box_g}
\Phi^3 \right) \left(\partial e^{-k \rho \Box_g}\Phi\right).
$$
The field equation reads
$$
(\xi^2 \Box_g+1)e^{-\Box_g}\Phi=V^{\prime}(\Phi),
$$
here and below we will denote by $\prime$ usual function derivative.
The Beltrami-Laplace operator takes the form
$\Box_g=-\pd_t^2-3H(t)\pd_t+\frac{1}{a^2}\pd_x^2$.

In the case of cubic potential the action (\ref{tach-ac})
corresponds to the the tachyon in bosonic String Field Theory \cite{Witten-SFT}
for the lowest level in the level truncated scheme\cite{Kost-Sam, West} with the induced
twisted space-time. It is supposed that that we are dealing
with $D3$-brane in the $D26$ space-time and the volume of the
$D22$ compactified subspace is omitted.

The case of quartic potential in the action (\ref{tach-ac})
corresponds to inclusion of the metric in the tachyon action of
the fermionic cubic string \cite{AMZ-PL} in the approximation of
a slowly varying auxiliary field \cite{AJK}. Such an action appears
 in the level truncated scheme where only lowest scalar
field (tachyon field) in the GSO(-) sector and
correspondingly the lowest scalar field in the GSO(+) sector
(this field does not have kinetic term and is considered as
auxiliary field) are considered. Integration over the auxiliary field results
in the quartic potential for the tachyon field. Note that
appearance of the non-extremal $D$-brane in the framework of
fermionic string field theory requires GSO(-) sector in the
spectrum \cite{BSZ,ABKM}.

The purpose of this work is to consider a general class of
nonlocal cosmological models of the form (\ref{tach-ac}) and
discuss which physical properties and consequences might be
obtained from such models. Among them we will be interested
firstly in the classical solutions of the corresponding
Friedmann equations which can be considered as a first
approximation to the quantum solutions and might be useful for
the study of ways to avoid the cosmological singularity
problem \cite{Turok}.

We start with the description of some general class of nonlocal
models. In section 3 we will describe new numerical algorithm
for solving nonlocal Friedmann equations. In section 4 we will
present the numerical solutions of Friedmann equations for
different parameters and will provide analytical explanation
for some of their interesting features. In section 5 we will
consider which cosmological solutions can be provided by
nonlocal models.

\section{General Class of Nonlocal Cosmological Models}

In this paper we will consider the following nonlocal scalar field
on the $D$-brane coupled to the gravity
\begin{equation}
\label{full-action}
S=\int d^4x\sqrt{-g}\left[
  \frac{M_p^2}{2}R +
  \frac{M_s^4}{g_4}\left(
      \frac{\xi^2}{2}\phi(\square_g/M_s^2)\phi +
      \frac{1}{2}\phi^2 -
      V(e^{k\square_g/M_s^2}\phi) -
      \Lambda^\prime-T \right)
  \right],
\end{equation}
where $g$ is the metric,
$\square_g=\frac1{\sqrt{-g}}\pd_{\mu}\sqrt{-g}g^{\mu\nu}\pd_{\nu}$,
$M_p$ is a Planck mass, $M_s$ is a characteristic string scale
related with the string tension $\alpha^{\prime}$ as
$M_s=1/\sqrt{\alpha^{\prime}}$, $\phi$ is a scalar field,
$g_4$ is a dimensionless four dimensional effective coupling constant related with the
ten dimensional string coupling constant $g_0$ and the
compactification scale, $\Lambda = \frac{M_s^4}{g_4}\Lambda^\prime $ is an
effective four dimensional cosmological constant, $T$ is the
brane tension, $k$ is a nonlocal coupling constant \cite{ABKM,AJK}.

For further investigation let us rewrite our action in
dimensionless space-time variables
\begin{equation}
\label{action}
S=\int d^4x\sqrt{-g}\left(\frac{m_p^2}{2}R+
\frac{\xi^2}{2}\phi \square_g \phi
+\frac{1}{2}\phi^2-V(\Phi)-\Lambda^\prime-T \right),
\end{equation}
where
$\phi$ is a dimensionless scalar field,
$\Phi=e^{k \square_g}\phi$ and $m_p^2=g_4 \frac{M_p^2}{M_s^2}$.

As a particular metric we take FRW one
\begin{equation}
\label{mFr}
ds^2={}-dt^2+a^2(t)\left(dx_1^2+dx_2^2+dx_3^2\right).
\end{equation}

We will consider spatially homogeneous configurations for which
Beltrami-Laplace operator takes the form
$\Box_g=-\partial_t^2-3H(t)\partial_t,$ for the convenience
of numerical calculations let us introduce
the following notation
$\Dh\equiv\partial_t^2+3H(t)\partial_t$,
which we will understand as generalization of second order time
derivative, which contains extra term $3H(t)\partial_t$
associated with gravitational background.

Equation of motion for the space homogeneous configurations for
the scalar field $\Phi$ takes the form
\begin{equation}
\label{eom-phi-H}
(-\xi^2\Dh+1)e^{2k \Dh}\Phi = V^{\prime}(\Phi).
\end{equation}

The Friedmann equations have the following form
\begin{equation}
\begin{split}
3H^2&=\frac{1}{m_p^2}~{\cal E},\\
3H^2+2\dot H&={}-\frac{1}{m_p^2}~{\cal P}.
\end{split}
\end{equation}

For the case of nonlocal potentials the energy  and the pressure
have additional non-local
terms ${\cal E}_{nl1}$ and ${\cal E}_{nl2}$
\begin{eqnarray}
\mathcal{E}&=
\underbrace{\mathcal{E}_k+\mathcal{E}_p+T+\Lambda^\prime}_{
  \text{\textit{usual local energy expression}}}
+~~\mathcal{E}_{nl1}~+~\mathcal{E}_{nl2},\\
\mathcal{P}&=
\underbrace{\mathcal{E}_k-\mathcal{E}_p-T-\Lambda^\prime}_{
  \text{\textit{usual local pressure expression}}}
-~~\mathcal{E}_{nl1}~+~\mathcal{E}_{nl2},
\end{eqnarray}
where
\begin{subequations}
\begin{equation}
\label{El-Enl}
 {\cal E}_k=\frac{\xi^2}{2}(\partial\phi)^2,
 \qquad {\cal E}_p={}-\frac{1}{2}\phi^2+V(\Phi),
\end{equation}
\begin{equation}
{\cal E}_{nl1}=k\int_{0}^{1} d \rho \left(e^{-k\rho \Dh}V^{\prime}(\Phi)\right)
 \left(\Dh e^{k\rho\Dh}\Phi\right),
\end{equation}
\begin{equation}
{\cal E}_{nl2}=-k\int_{0}^{1}  d \rho\left(\partial e^{-k \rho\Dh}
V^{\prime}(\Phi)\right) \left(\partial e^{k \rho \Dh}\Phi\right).
\end{equation}
\end{subequations}

As we can see the structure of Friedmann equations and of
equation for scalar field in this model is rather complicated
and their study is already a very interesting mathematical
problem and can be considered as a separate mathematical
investigation.

To avoid calculation of $e^{-k \rho\Dh}$ term which
is much harder to compute than $e^{k \rho\Dh}$
($k>0$) as computation of the former results in an ill-posed
problem \cite{LY} we will use the following representation for
nonlocal energy terms on the equation of motion for the scalar
field
\begin{subequations}
\begin{equation}
\label{Enl-goodrepr}
{\cal E}_{nl1}=k\int_{0}^{1}d \rho \left((-\xi^2\Dh+1)
e^{(2-\rho)k \Dh}\Phi \right)
 \left(\Dh e^{k\rho\Dh}\Phi\right),
\end{equation}
\begin{equation}
{\cal E}_{nl2}=-k\int_{0}^{1} d \rho \left(\partial (-\xi^2\Dh+1)
e^{(2-\rho)k\Dh}
\Phi\right) \left(\partial e^{k \rho \Dh}\Phi\right).
\end{equation}
\end{subequations}

\section{Method for Solution of Nonlocal Friedmann\\ Equations}

\subsection{Iterative Procedure for Solution Construction}
\label{ITforSS}
For numerical calculations let us rewrite our system in the following form
\begin{subequations}
\label{fr}
\begin{equation}
\label{fr1}
3H^2=\frac{1}{m_p^2}~{\cal E},
\end{equation}
\begin{equation}
\label{fr2}
\dot H={}-\frac{1}{2m_p^2}~(\cal P+\cal E),
\end{equation}
\begin{equation}
\label{field-eq}
(-\xi^2 \Dh+1)e^{2k\Dh}\Phi =V^{\prime}(\Phi).
\end{equation}
\end{subequations}

In this paper we would like to point out which physical results
we can extract from solutions of this system.  First, we will
find Hubble function and the scalar field from equations
(\ref{fr2}) and (\ref{field-eq}). It happens that equation
(\ref{fr1}) plays a role of energy conservation from which the
effective cosmological constant is extracted and is unique for
every field configuration.

Integrating equation (\ref{fr2}) we get the following system
\begin{subequations}
\label{eqns}
\begin{equation}
(-\xi^2 \Dh+1)e^{2k\Dh}\Phi =V^{\prime}(\Phi),
\end{equation}
\begin{equation}
\label{fr2-proint}
H=-\frac{1}{m_p^2} \int_0^t d \tau ~\left[\frac{\xi^2}{2}(\partial\phi)^2
-k\int_{0}^{1} d \rho \left(\partial (-\xi^2 \Dh+1) e^{(2-\rho)k\Dh}
\Phi\right) \left(\partial e^{k \rho \Dh}\Phi\right)\right].
\end{equation}
\end{subequations}

To find solutions of the system (\ref{eqns}) we construct the following iterative process
\begin{subequations}
\label{iter-procedure}
\begin{equation}
\Phi = \lim_{n\to\infty} \Phi_n, ~~H = \lim_{n\to\infty} H_n,
\end{equation}
where iterations $\Phi_n$ and $H_n$ are in turn obtained as limits
of sub-iterations
\begin{equation}
\Phi_{n+1} = \lim_{m\to\infty} \Phi_{n,m}, ~~H_{n+1} = \lim_{m\to\infty} H_{n,m},
\end{equation}
which are recursively defined as ($m \geqslant 0$)
\begin{eqnarray}
\label{it-pr-onefieldforgivenH}
\Phi_{n,m+1}&=&{V^{\prime}}^{-1}\left( (-\xi^2\DD{n}+1)e^{2k\DD{n}}\Phi_{n,m}\right),\\
\label{fr2-proint-itpr}
H_{n,m+1}&=&-\frac{1}{m_p^2}\int_0^t d\tau
\left[
  \frac{\xi^2}{2}(\pd e^{k\DD{n,m}}\Phi_{n+1})^2-
\right.
  \\&&\notag
\left.
  k\int_{0}^{1}d\rho
  \left(\partial (-\xi^2 \DD{n,m}+1) e^{(2-\rho)k\DD{n,m}}\Phi_{n+1}\right)
  \left(\pd e^{k \rho \DD{n,m}}\Phi_{n+1}\right)
\right],
\end{eqnarray}
where initial iteration in $m$ is taken as
\begin{equation}
\Phi_{n,0}=\Phi_n, ~~H_{n,0}=H_n.
\end{equation}
\end{subequations}
By ${V^\prime}^{-1}$ we denote a function inverse to $V^\prime$.
Note how sub-iteration $\Phi_{n,m}$ depends only on $H_n$, i.e. the whole step of
sub-iterations for $H$. Note also that $H_{n,m}$ depends on $\Phi_{n+1}$, this
is essentially a way to accelerate convergence.

Intuitively this iteration procedure does the following. First
fixing $H_n$ it finds $\Phi_{n+1}$ using iteration process
(\ref{it-pr-onefieldforgivenH}) which is a natural
generalization of a process which is described in
\cite{BFOW,MolZw,Yar-JPA} and for which convergence is proved
analytically. Then $H_{n+1}$ is obtained by iterating
(\ref{fr2-proint-itpr}) where $\Phi_{n+1}$ is fixed as computed
in the previous step. Thus found $H_{n+1}$ is used to find
$\Phi_{n+2}$ and so on. This kind of multi-step iterative
procedure are known to provide rapid convergence. In our
experiments this iterative procedure converges much more
reliably as compared to more naive iterative procedure where
one does not do intermediate iterations.

In a similar fashion one can construct iterative procedures for
systems with several scalar fields such as those studied in
\cite{LY,Vernov-twofields}.

\subsection{Exponentiating $\Dh$ by Solving Diffusion Equation}

Equations of motion as well as the iteration method discussed
above require computation of action with operator
$e^{k \rho\Dh}$ on a given function. Computationally this is a
nontrivial problem by itself. We use the following method which
beside giving mathematical definition for such an operator more
importantly provides corresponding computational method.

Result of the acting with operator $e^{\rho \Dh}$ on a given
function $\varphi(t)$ is natural to define as a solution of the
following diffusion partial differential equation with
corresponding initial and boundary conditions
\begin{eqnarray}
\label{f-rho-t}
\frac{\pd}{\pd \rho} f(\rho, t) & = & \Dh f(\rho, t),\\
f(0, t) & = & \varphi(t),\\
\label{f-rho-boundary}
f(\rho, \pm \infty) & = & \varphi(\pm \infty),
\end{eqnarray}
where $-\infty < t < +\infty$, $\rho \geqslant 0$. Note that here
evolution goes along $\rho$ axis, while $t$ plays a role of space.
Solution of (\ref{f-rho-t}) is a rigorous mathematical definition of
the operator $e^{\rho \Dh}$ acting on function $\varphi$
$$
e^{\rho\Dh}\varphi(t) = f(\rho,t).
$$
Such solution exists and is unique for wide classes of $H$ and
$\varphi$ \cite{Y-lic}. Since in standard settings such a
diffusion problem is well defined in particular for
$H,\varphi\in\mathfrak{L}_2(\mathbb{R})$ which seem to suffice
applications which are considered here we will not discuss this
topic any further.

\subsection{Fixing Potential and Other Parameters}

So far we did not specify exact shape of the potential for our
system. In this paper we are interested in coupled to the gravity
a tachyon field which appears as a lowest excitation in level
truncation scheme for fermionic string field theory \cite{AJK}.
Thus all further discussion will be about quartic potential
\begin{equation}
\label{potential}
V(\Phi)=\frac{1}{4}\Phi^4.
\end{equation}
We would like to note though that
from numerical perspective this potential can lead to
difficulties as it results in solutions which are not
differentiable at zero (they behave as $t^{1/3}$ at in the
vicinity of zero \cite{Vlad}). For the later reason we
found it useful to study similar problem in potential
$$
V_\alpha(\Phi)=(1-\alpha) \frac{1}{2}\Phi^2 + \alpha\frac{1}{4}\Phi^4,~~0 < \alpha < 1.
$$
As $\alpha\to 1$ this potential tends to potential (\ref{potential}).
The interesting property of this potential is that solution has
finite derivative at zero and moreover there is a theorem which
states convergence of our type of iterative procedure in case
of Minkowski metric \cite{LJ}.

Let us return to the potential (\ref{potential}).
Equation (\ref{field-eq}) is invariant under shifts
\begin{eqnarray}
\Phi(t)\rightarrow \Phi(t+t_0),~~H(t) \rightarrow H(t+t_0),
\end{eqnarray}
and mirroring
\begin{eqnarray}
\Phi(t)\rightarrow -\Phi(-t),~~H(t) \rightarrow -H(-t).
\end{eqnarray}
Thus it is natural to look for odd solutions. Without loss of
generality we can set $\Phi(0)=0$.

Equation (\ref{field-eq}) has the following constant solutions
$$
\Phi=0, \pm 1.
$$
Following the same logic as in Minkowsky case \cite{MolZw,AJK} since
$V(\Phi=+1)=V(\Phi=-1)$ we can expect existence of kink-type solutions
where scalar field interpolates between $\Phi=+1$ and $\Phi=-1$.

Iterative procedure (\ref{iter-procedure}) requires initial iteration
$\Phi_0$, $H_0$ to be specified explicitly.
To find such kink solutions it is natural to take
$$
\Phi_0(t)=\sign(t),~~H_0(t)=0.
$$
This initial iteration leads to a rapid convergence. It is
interesting that our numerical experiments show that iterative
procedure (\ref{iter-procedure}) converges to a kink-type
solution if we start from any two bounded functions which are
positive on positive semi-axis and negative on negative one.
This observation provides a strong indication for uniqueness of
the solutions (up to shifts and mirroring).

The value of $\xi^2$ is determined during level-truncation
procedure, its value is
\begin{equation}
\label{xi2-phys}
\xip=-\frac{1}{4\log{\frac{4}{3\sqrt{3}}}}\approx 0.96.
\end{equation}
Nevertheless since different values of $\xi^2$ already in
Minkowski case result in different physical pictures we will systematically consider various
values of $\xi^2$. Note also that the case $\xi^2=0$ corresponds
to $p$-adic string with $p=3$.
The value of nonlocal coupling constant is fixed by the same procedure
it is $k=\frac{1}{8}$ in our case.

\section{Numerical Results}

\subsection{Numerical Solutions for Friedmann Equations}
\label{NSforFE}

\begin{figure}
\centering
\includegraphics[width=41mm]{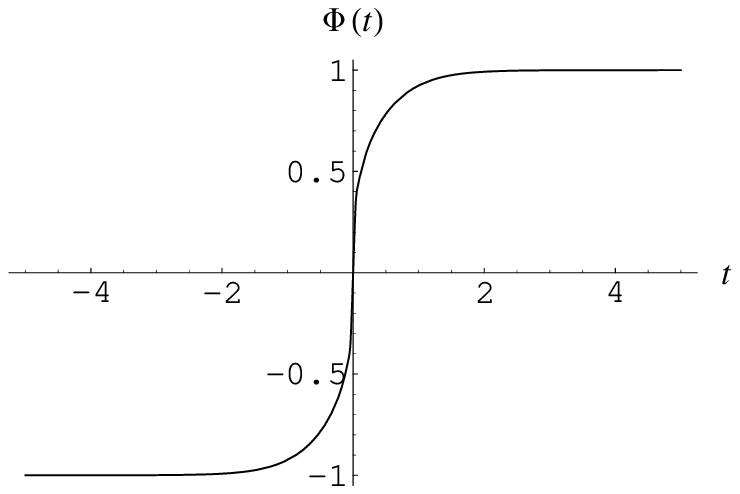}
\includegraphics[width=41mm]{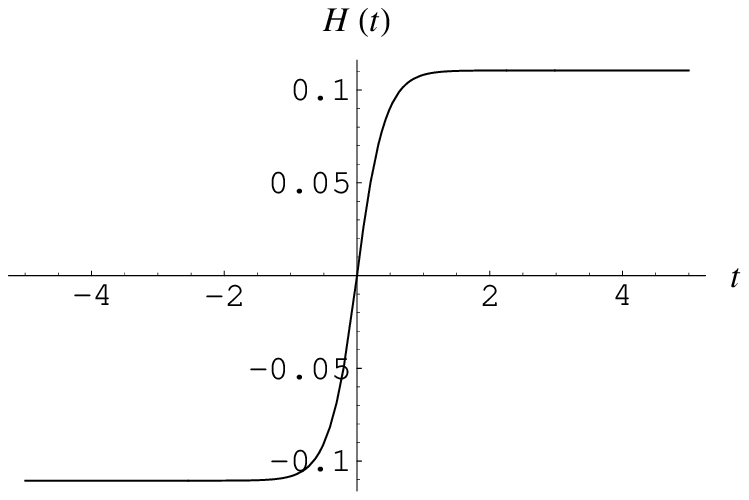}
\includegraphics[width=41mm]{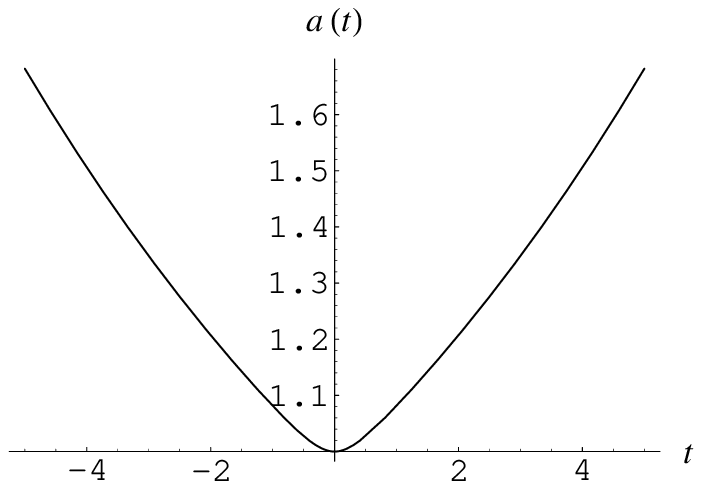}\\
\includegraphics[width=41mm]{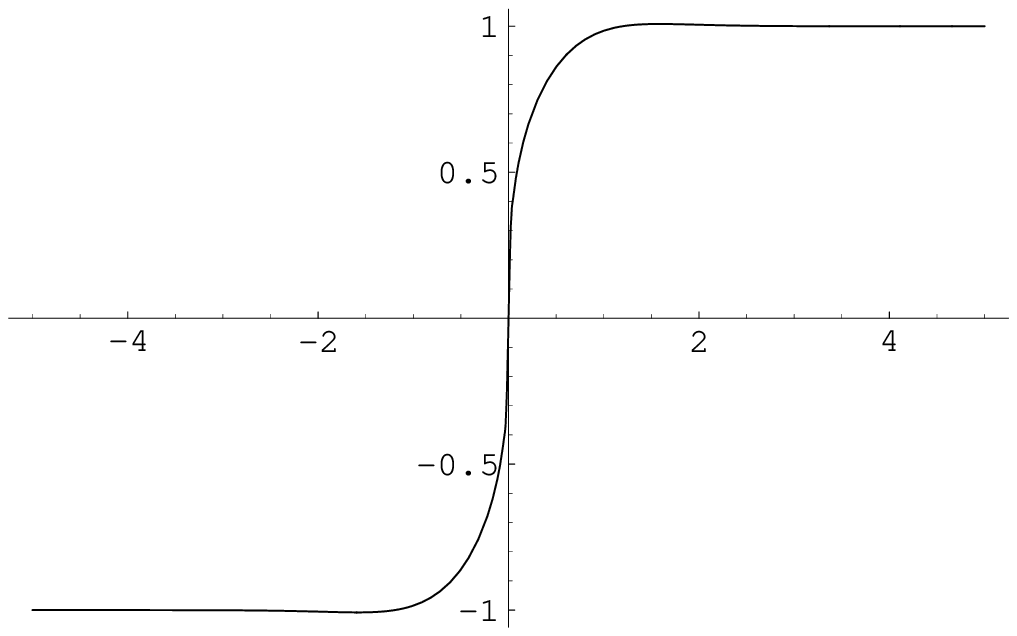}
\includegraphics[width=41mm]{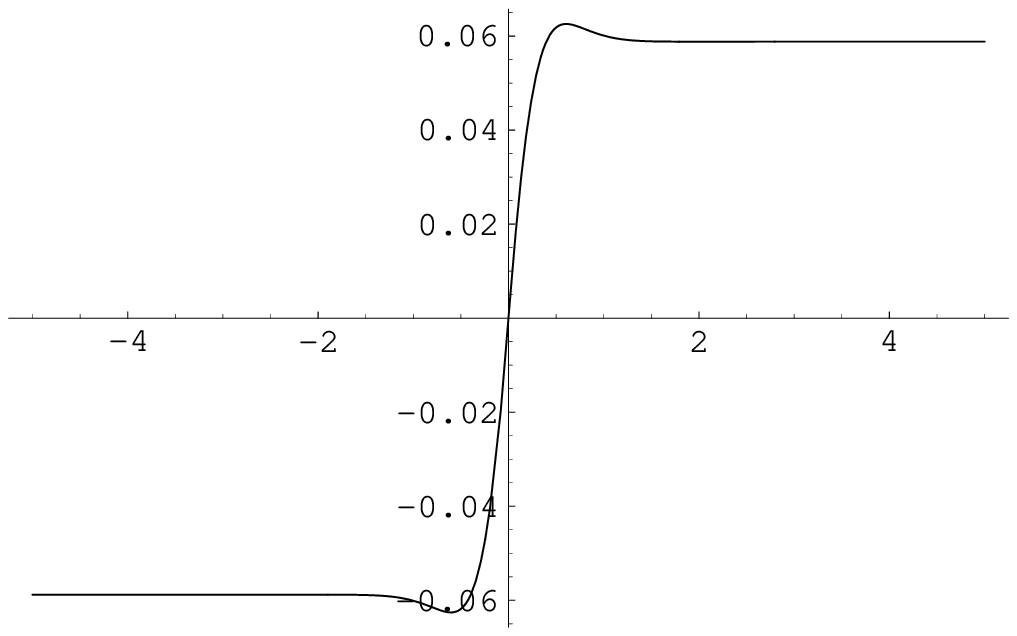}
\includegraphics[width=41mm]{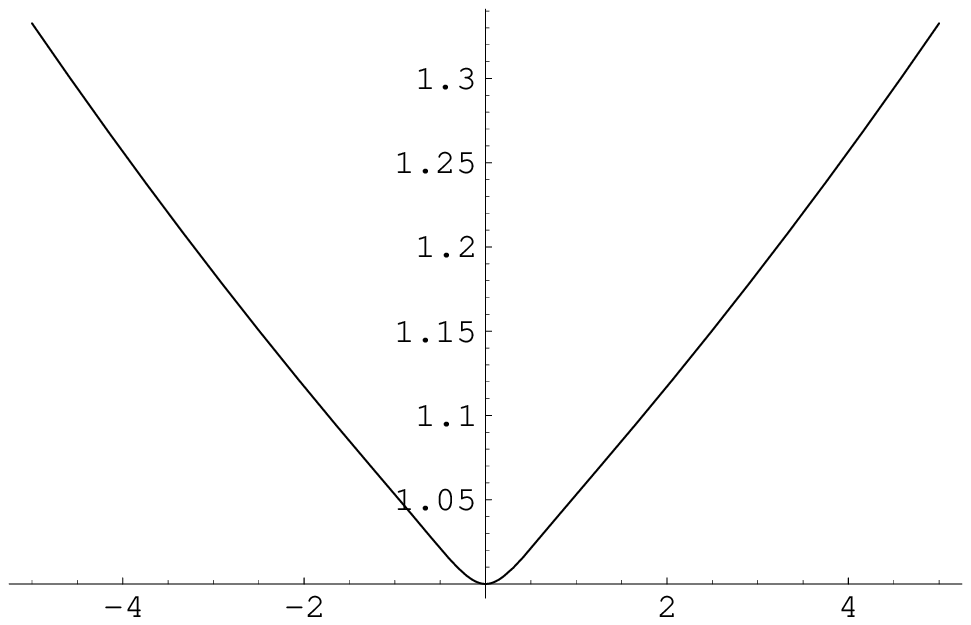}\\
\includegraphics[width=41mm]{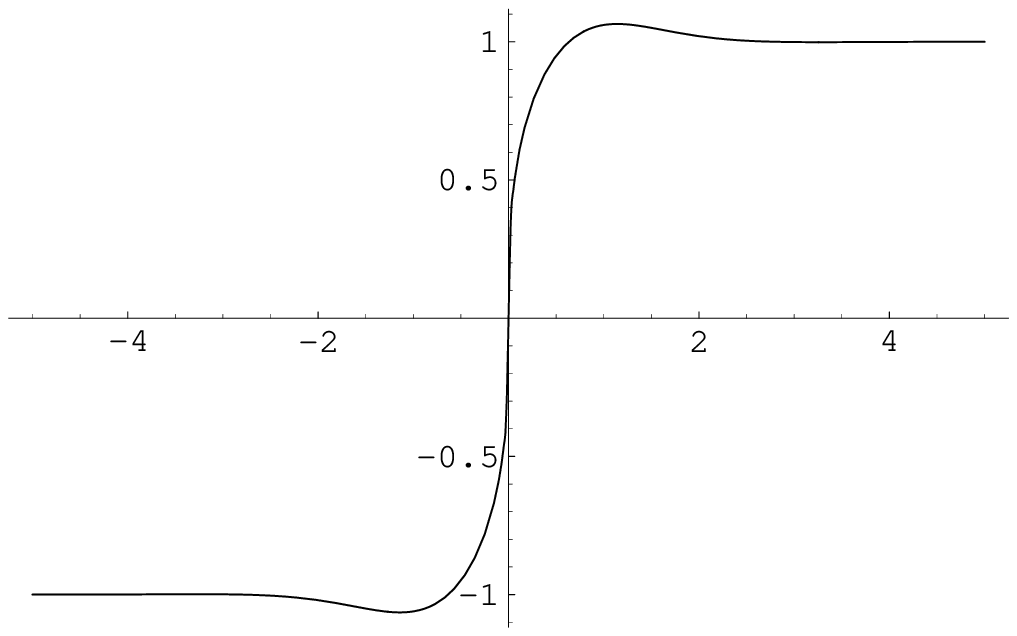}
\includegraphics[width=41mm]{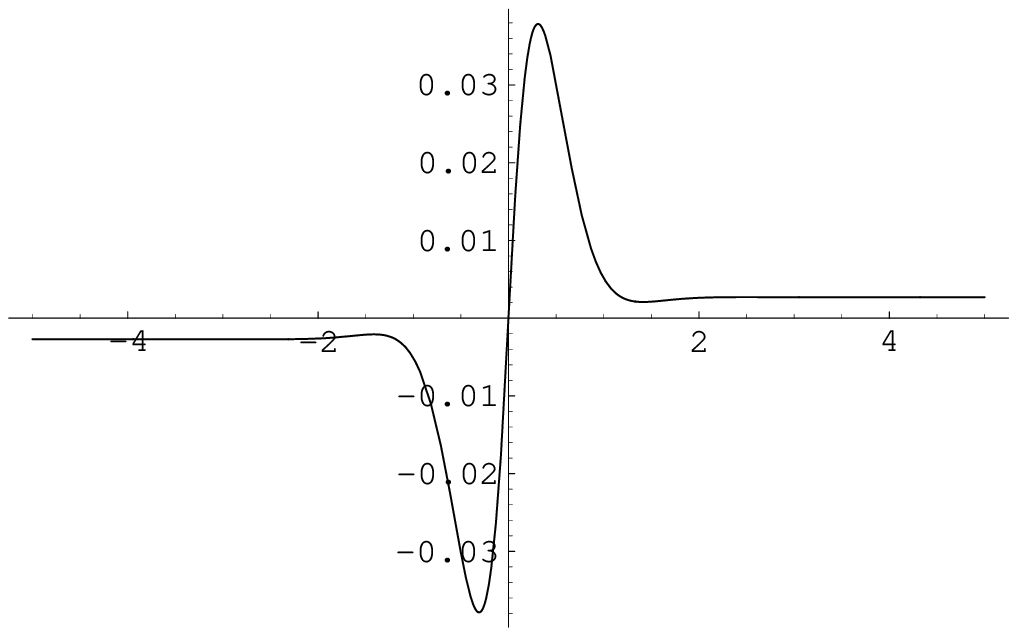}
\includegraphics[width=41mm]{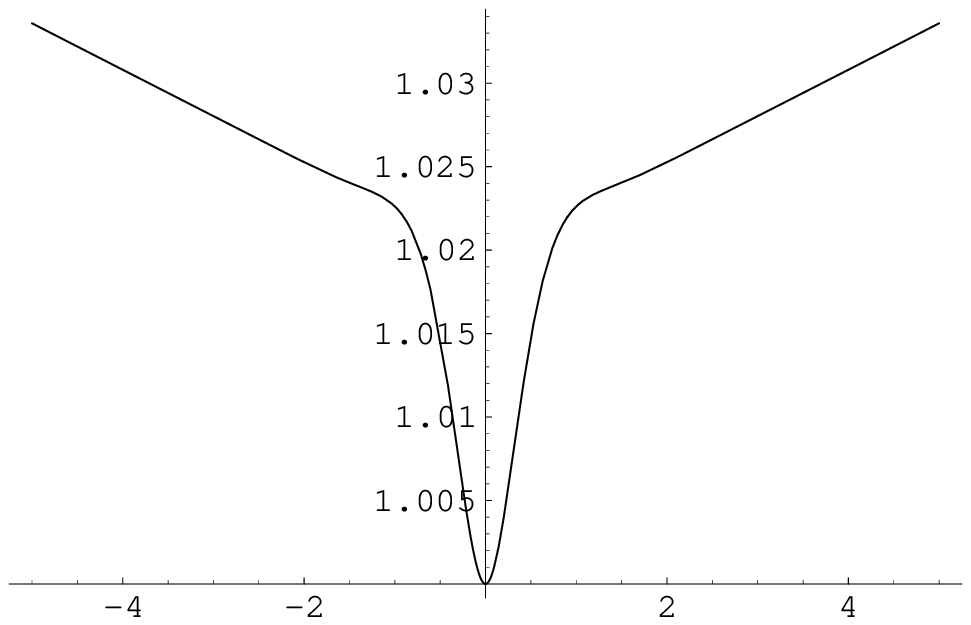}\\
\includegraphics[width=41mm]{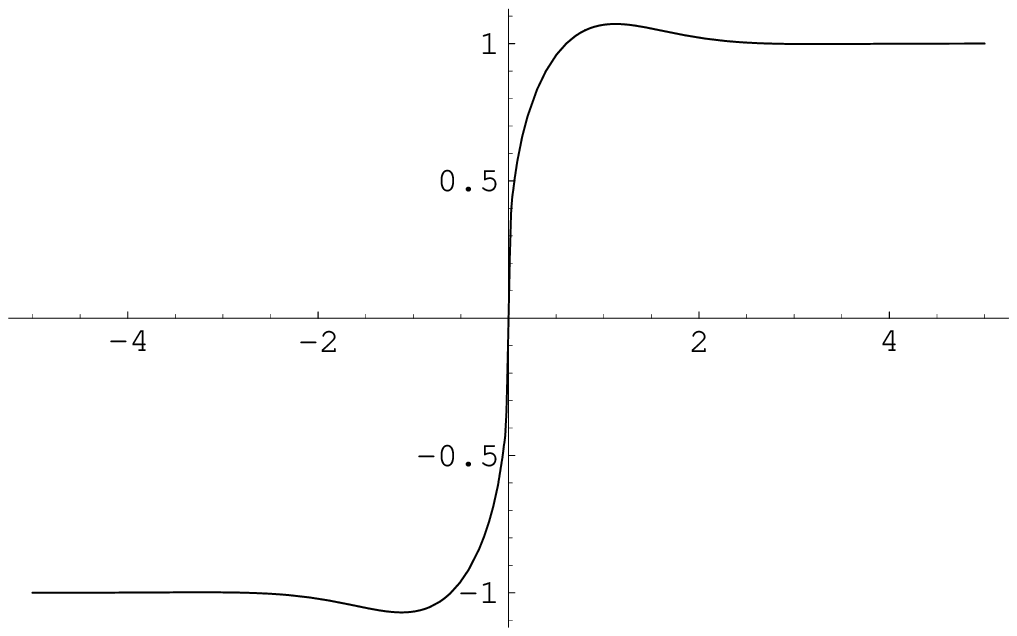}
\includegraphics[width=41mm]{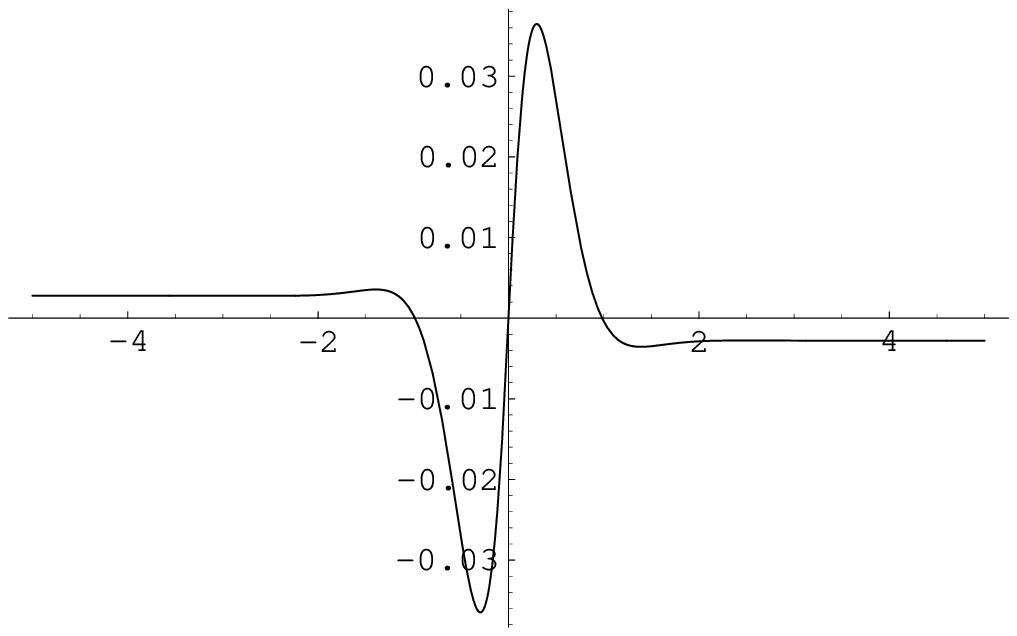}
\includegraphics[width=41mm]{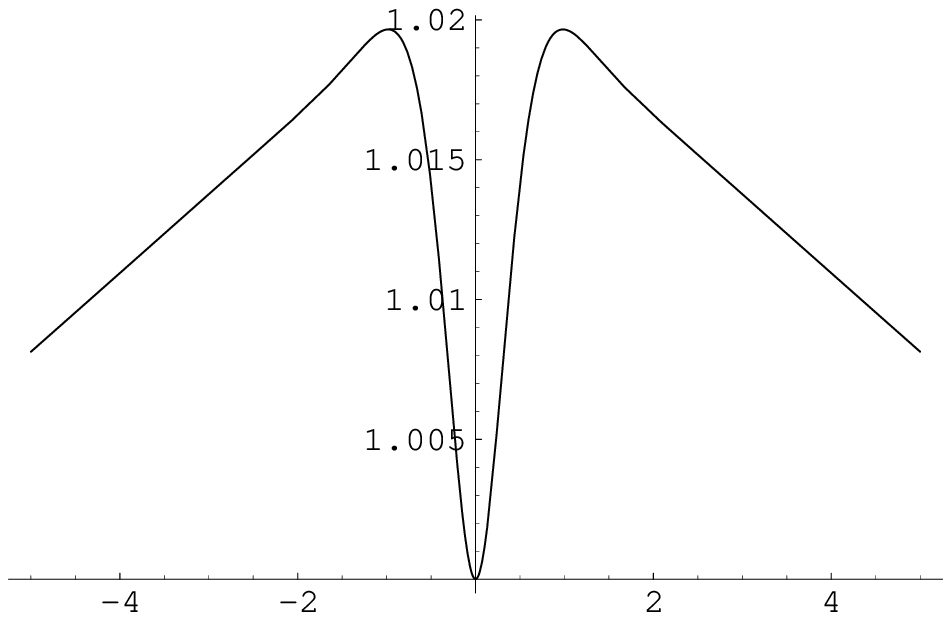}\\
\includegraphics[width=41mm]{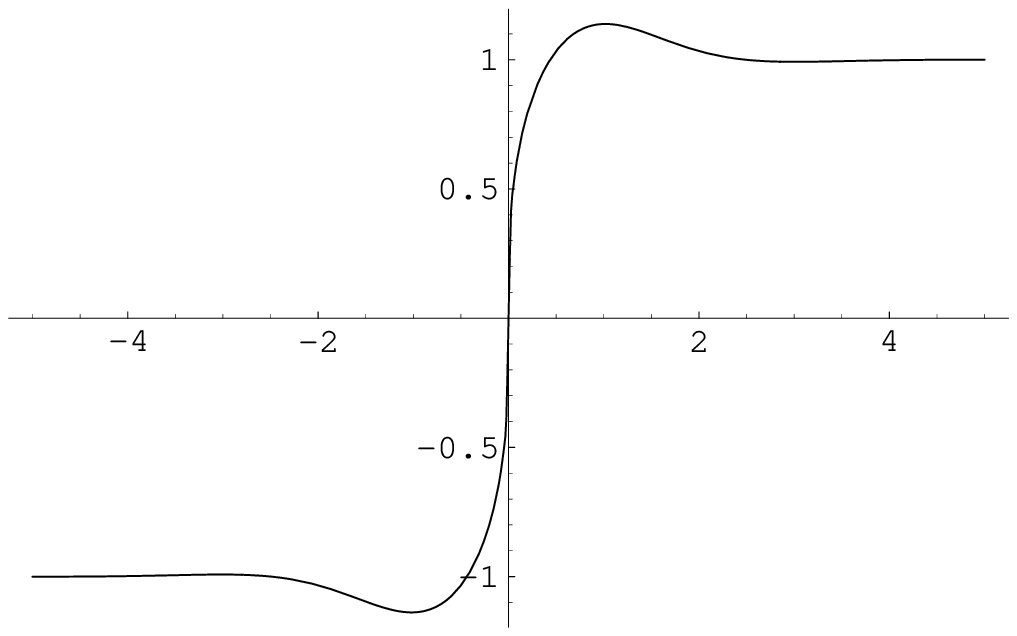}
\includegraphics[width=41mm]{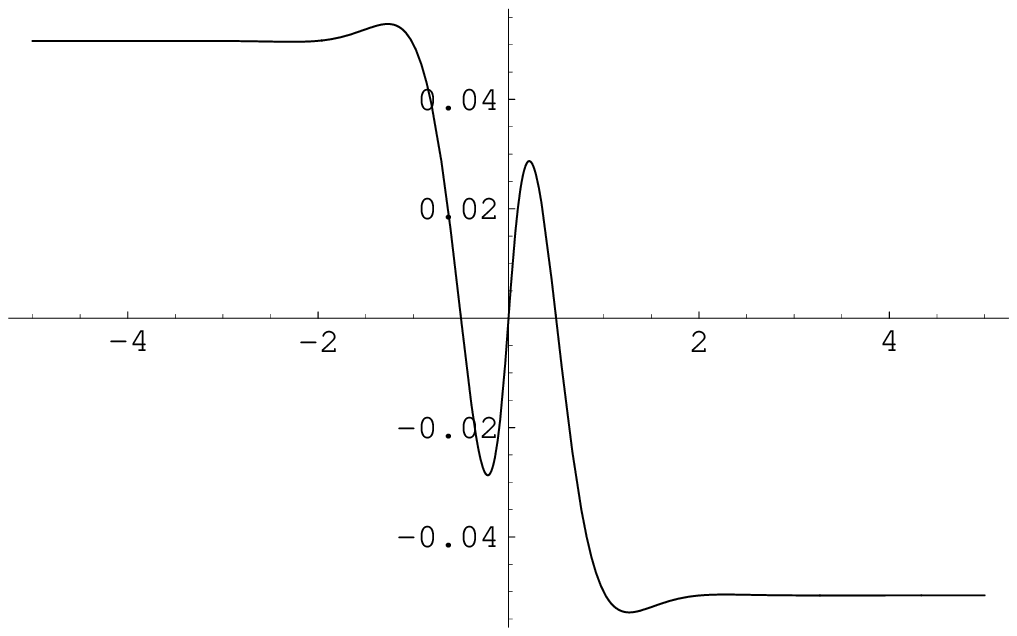}
\includegraphics[width=41mm]{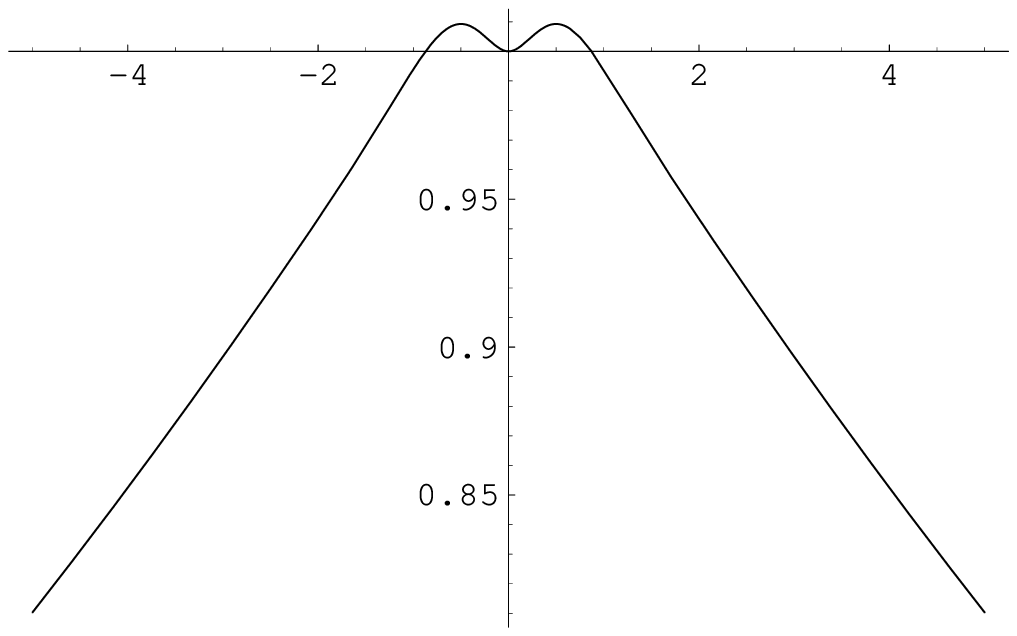}\\
\includegraphics[width=41mm]{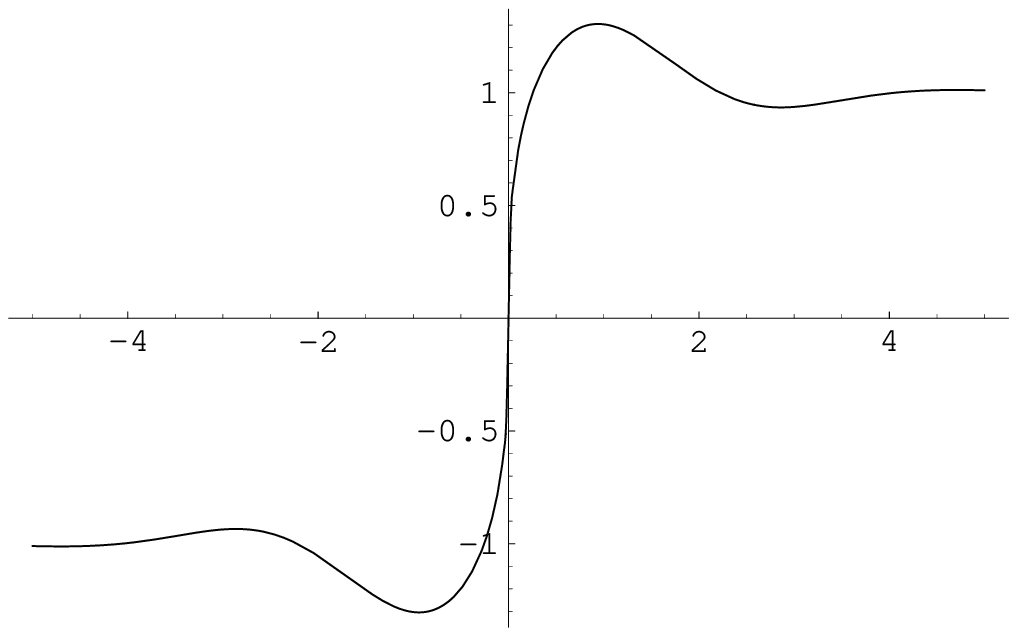}
\includegraphics[width=41mm]{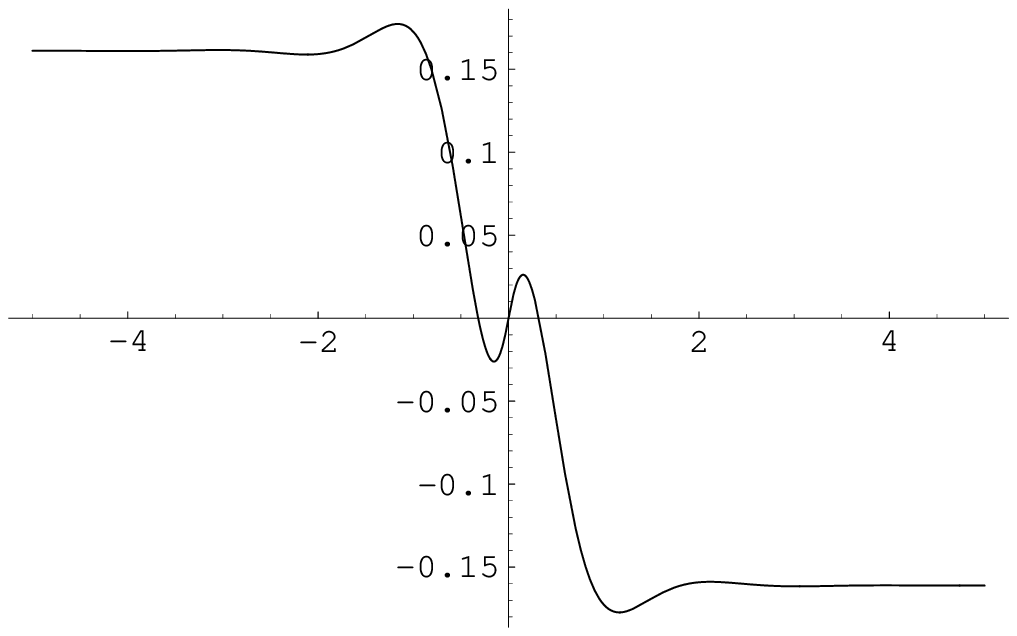}
\includegraphics[width=41mm]{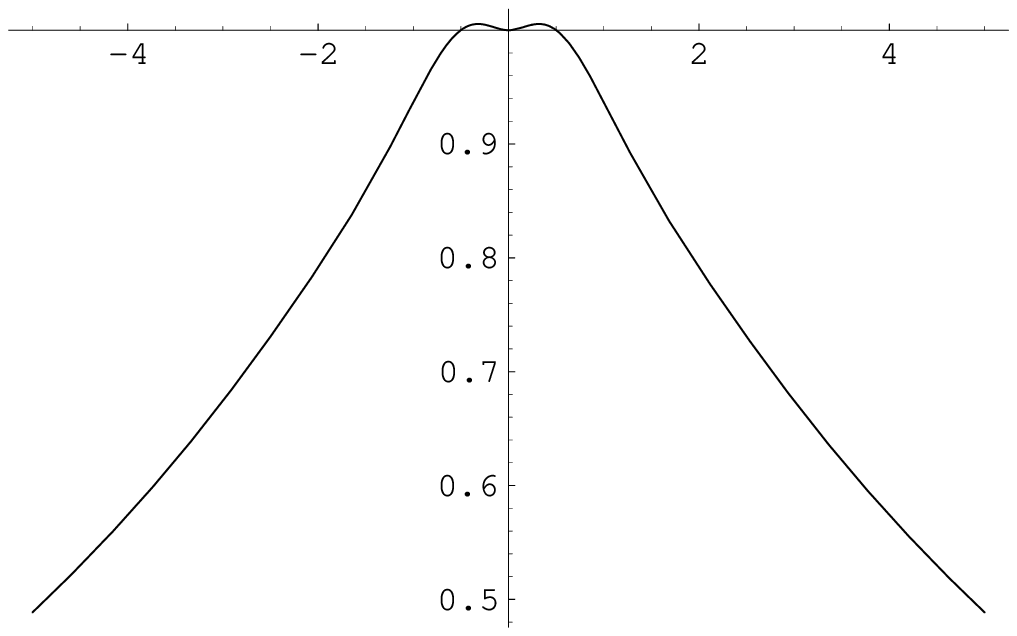}\\
\includegraphics[width=41mm]{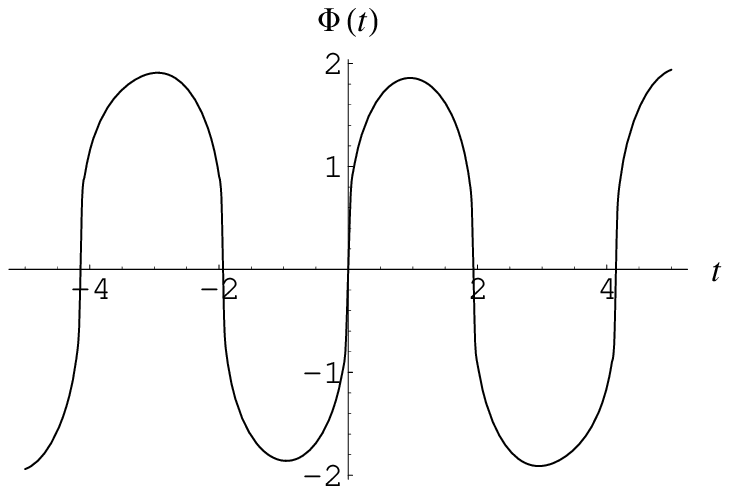}
\includegraphics[width=41mm]{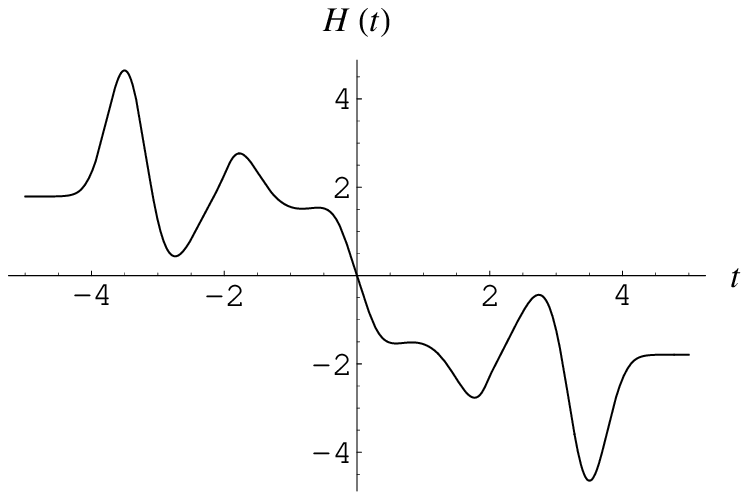}
\includegraphics[width=41mm]{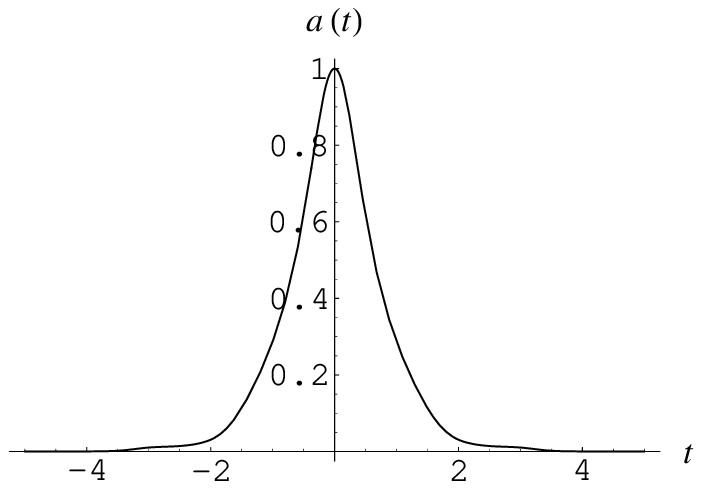}
\caption{Solutions of the Friedmann equations $\Phi$, $H$
and $a$ (left to right) for different values of the parameter
$\xi^2=0,0.2,0.41,0.43,0.6, 0.96, 2$ (top to bottom). Note that
on the figures for scale factor $a(t)$ axes cross at the point $(t=0, a=1)$.
We can see that when $\xi^2$ increases the shape of the scale factor changes
from parabolic type ($a(t)\geqslant 1$ for all times)
to lump type ($a(t) \leqslant 1$ for all times);
transition between these shape types is illustrated on the Fig. \ref{xi-shape}.}
\label{H-iter}
\end{figure}

\begin{figure}
\centering
\includegraphics[width=41mm]{phi-xi2=0.41.eps}
\includegraphics[width=41mm]{H-xi2=0.41.eps}
\includegraphics[width=41mm]{a-xi2=0.41.eps}\\
\includegraphics[width=41mm]{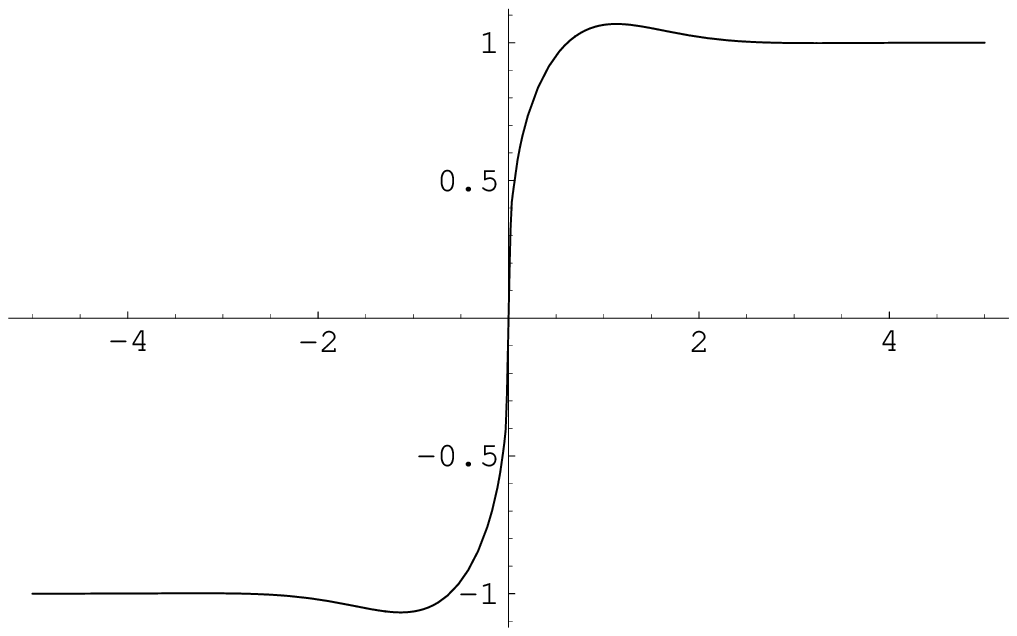}
\includegraphics[width=41mm]{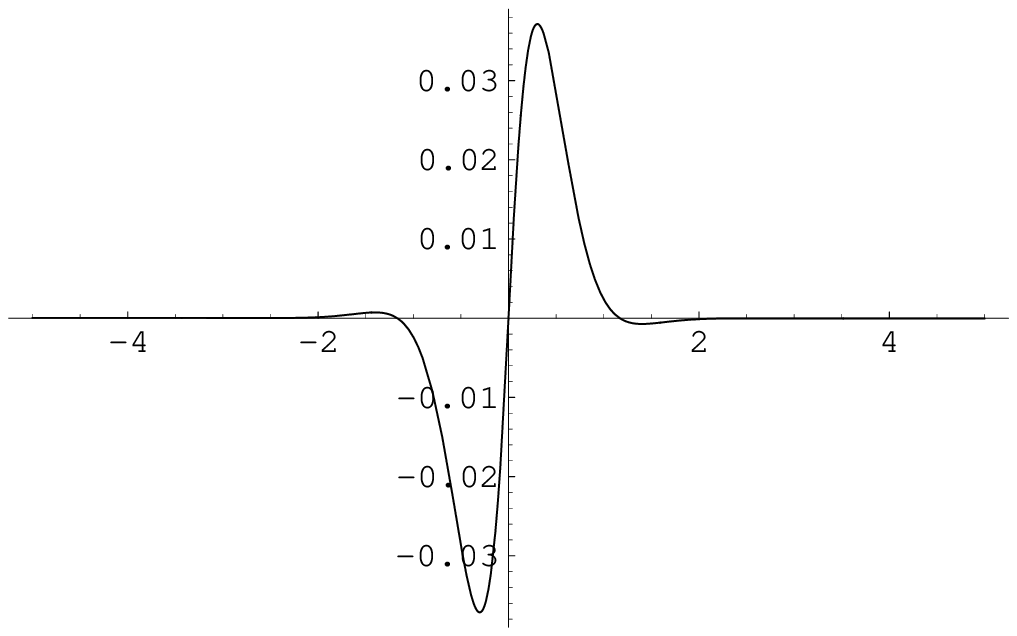}
\includegraphics[width=41mm]{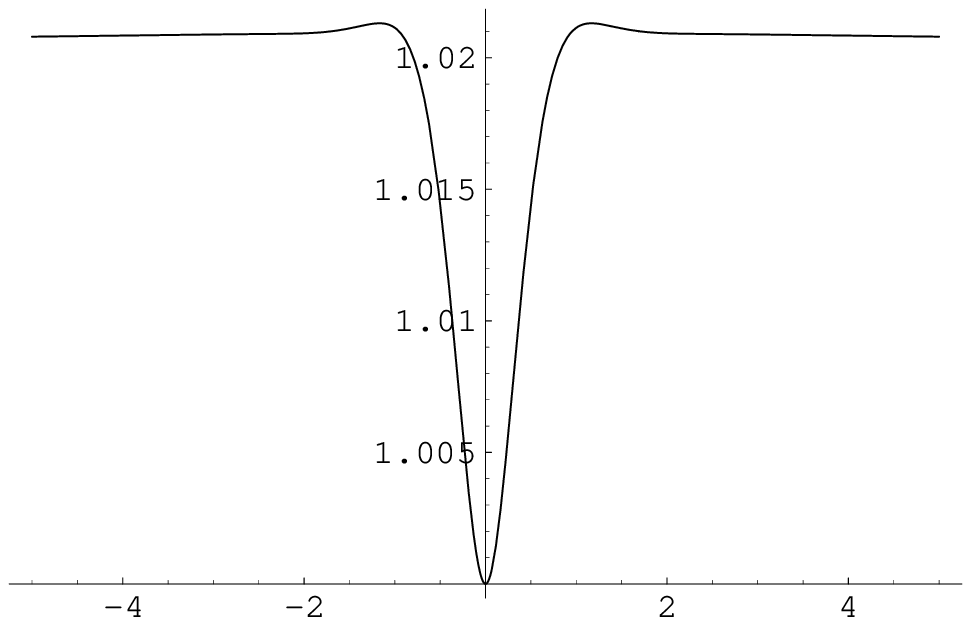}\\
\includegraphics[width=41mm]{phi-xi2=0.43.eps}
\includegraphics[width=41mm]{H-xi2=0.43.eps}
\includegraphics[width=41mm]{a-xi2=0.43.eps}
\caption{Functions $\Phi$, $H$ and $a$ (left to right) for $\xi^2=0.41,0.42,0.43$
(top to bottom). Note how asymptotic behavior changes as we go through
$\xi^2=0.42$.}
\label{xi-shape}
\end{figure}

In what follows for numerical computations we put $m_p=1$. We
used iterative method presented in the previous section to
solve Friedmann equations for our system. We observe the crucial
role of the value of $\xi^2$, as it varies the following
physical properties of the system are affected (see Fig.\ref{H-iter}).
\begin{itemize}
\item
For $\xi^2=0$ the solution in both $\Phi$ and $H$ has a
monotonic kink shape. This behavior is qualitatively
similar to the case of Minkowsky metric
\cite{AJK,Yar-JPA,VladVol}.

\item
There exists a critical value
$\xio\approx 1.18$
which determines between types of the $\Phi$ component of the
solution -- kink or oscillatory with finite period. For
$\xi^2 < \xio$
the solution has a kink shape with exponentially
decreasing oscillations around $\pm 1$ as $t\to\pm\infty$.
On the other hand if
$\xi^2 > \xio$
the solution converts to oscillations with finite period.\footnote{
Note that for iterative method which we present in this paper it is essential
that solution $\Phi$ has well defined limits at infinite times.
This fact is used in particular for the boundary condition
(\ref{f-rho-boundary}). This means that this method is not
suitable for finding oscillatory solutions with high precision.
Oscillatory behavior reported in this paper was obtained by
vastly enlarging the lattice in $t$, this did not allow finding
critical value of $\xi^2$ with more then $3$ significant digits.
Reliable methods for finding oscillatory solutions for the type
of equations considered in this paper is an open mathematical problem even for
Minkowski metric, see \cite{Yar-JPA}.
Note though that physically significant solutions in this model turned out to be of kink
type where iterations rapidly converge.}
Similar behavior was observed in the case of Minkowsky
metric, although in that case the critical value was higher
$\approx 1.38$ \cite{AJK,Yar-JPA}.

\item
There is one more critical value
$\xis\approx 0.42$ which determines a shape
of a Hubble function (see Fig.\ref{xi-shape}). More precisely,
as $\xi^2$ grows Hubble function $H$ gets a turning point at
some positive time $t_0$. Moreover, for $\xi^2 > \xis$ $H$ ends
up in tending a negative (positive) value as
$t\to\infty$ ($t\to-\infty$). This is a new type of
behavior which was not present in Minkowsky case.

Note that the physical value of $\xi^2$ (\ref{xi2-phys})
is in the region
$$
\xis < \xip < \xio.
$$
\end{itemize}

Figures \ref{H-iter}, \ref{xi-shape} also show how dynamics
of the scale factor $a(t)$
\begin{equation}
\label{scale-factor}
a(t)=a_0 \exp\left(\int_0^t H(\tau)d\tau\right)
\end{equation}
changes as $\xi^2$ increases. For $\xi^2=0$ it has a has a
minimum at the perturbative vacuum $\phi=0$ and increases as
$\phi$ tends to nonperturbative vacua $\pm 1$. As $\xi^2$
increases the shape of $a$ changes and for $\xi^2 > \xis$ the
scale factor decreases as $\phi\to\pm 1$ ($t\to\pm\infty$). For
numerical computations we use initial condition $a_0=1$ in
(\ref{scale-factor}).

\subsection{Two Profiles of the Hubble Function}

To analyze the late time behavior of Hubble function let us use
the so called mechanical approximation which was studied for
example in \cite{AJ-JHEP} and
gives realistic qualitative picture at least for the late time
behavior. Mechanical approximation is constructed by keeping in
original nonlocal expression only terms with derivatives no
higher than second order. For (\ref{fr2-proint}) we get
\begin{equation}
H(t)\approx
-\frac{1}{m_p^2}\int_0^t d\tau (\frac{\xi^2}{2}(\pd \phi)^2-k(\pd \phi)^2)
=\frac{1}{m_p^2}\int_0^t d\tau (k-\frac{\xi^2}{2})(\pd \phi)^2.
\end{equation}
We can see from the expression above that there exists a value
of $\xi^2$ which determines the change of sign for the Hubble
function, more precisely for $\xi^2 < 2k$ the Hubble function
is positive for $t\to\infty$ (negative as $t\to-\infty$), while
for $\xi^2 > 2k$ the reverse becomes true. Thus mechanical
approximation gives us the critical value of $\xi^2=2k=0.25$.
As already described numerical computations give us the value
$\xis\approx 0.42$ which slightly differs from the just found
approximate critical $\xi^2$. This difference manifests the
influence of higher order nonlocal terms. Nevertheless the
analysis above clarifies the phenomenon which causes the
change of sign for the asymptotic values of the Hubble
function.

\subsection{Two Regimes of the Solution}

In this section we will address the mechanism which forms
numerically found exponentially decreasing oscillations and the
existence of critical value $\xio$. We will also compare this
behavior with results obtained without gravitational term in
action \cite{Yar-JPA}.

Let us present solution of equation (\ref{field-eq}) for the quartic potential
$V(\Phi)=\frac{1}{4}\Phi^4$
\begin{equation}
\label{field-eq-quartpot}
(-\xi^2 \Dh+1)e^{2k\Dh}\Phi=\Phi^3
\end{equation}
as a sum
\begin{equation}
\Phi(t)=\Phi_0(t)+\chi(t),
\label{dist-sol}
\end{equation}
where $\Phi_0$ denotes solution of equation (\ref{field-eq})
for $\xi^2=0$. Substituting (\ref{dist-sol}) to (\ref{field-eq})
and leaving only linear terms in $\chi$ we get
\begin{equation}
(-\xi^2 \Dh+1)e^{2 k \Dh}(\Phi_0+\chi)=\Phi_0^3+3\Phi_0^2\chi.
\end{equation}
Using the fact that $\Phi_0$ is the solution of equation (\ref{field-eq})
for $\xi^2=0$ we can write equation for $\chi(t)$ which in
linear approximation has the form
\begin{equation}
(-\xi^2 \Dh+1)e^{2 k \Dh}\chi=3\Phi_0^2\chi+\xi^2 \Dh e^{2 k \Dh}\Phi_0.
\end{equation}
Using the fact that operator $e^{2 k \Dh}$ acts as an identity on constants
we can write large $t$ approximation
\begin{equation}
(-\xi^2 \Dh+1)e^{2 k \Dh}\chi=3\chi,
\label{chi-eq}
\end{equation}
were we used asymptotical properties of solution $\Phi_0$.

To carry out harmonic analysis for equation (\ref{chi-eq})
we will consider eigenfunctions of the D'Alembertian operator
$\Dh\upsilon_\lambda=-\lambda \upsilon_\lambda$.
Using that $e^{2k\Dh}\upsilon_\lambda=e^{-2 k \lambda}\upsilon_\lambda$ and
expanding $\chi$ in $\upsilon_\lambda$ equation (\ref{chi-eq})
leads us to
\begin{equation}
\label{charact-eq}
(\xi^2 \lambda+1)e^{-2 k \lambda}\upsilon_\lambda=3\upsilon_\lambda.
\end{equation}
Our goal is to analyze which values of $\xi^2$ allow
real-valued $\lambda$, i.e. non-damping oscillations take place.
Considering equation (\ref{charact-eq}) as an equation for complex
variable $\lambda$ with parameter $\xi^2$ we obtain
that there is a minimum value of $\xi_0^2\approx 1.77$
for which $\lambda$ is real. So for $\xi^2<\xi_0^2$
we have solutions with nonzero imaginary parts which result in
vanishing solutions while $\xi^2>\xi_0^2$ leads to oscillatory regime.
Equation (\ref{charact-eq}) has exactly the same form in case
of Minkowsky space which was considered earlier
\cite{Yar-JPA} (functions $\upsilon_\lambda$ are different as they depend on $H$).
As we can see method discussed in this section provides us only qualitative
explanation of the changing of the regimes, the value of critical $\xi^2$ found here
only approximately reproduces numerically obtained $\xio\approx1.18$.
It is interesting to note that Minkowski case leads to higher value of
critical $\xi^2$ \cite{Yar-JPA}, i.e. it appears that ``friction'' $H(t)$ does not
damp oscillations but au contraire increases them.

\section{Cosmological properties}

\begin{figure}[ht!]
\centering
\includegraphics[width=43mm]{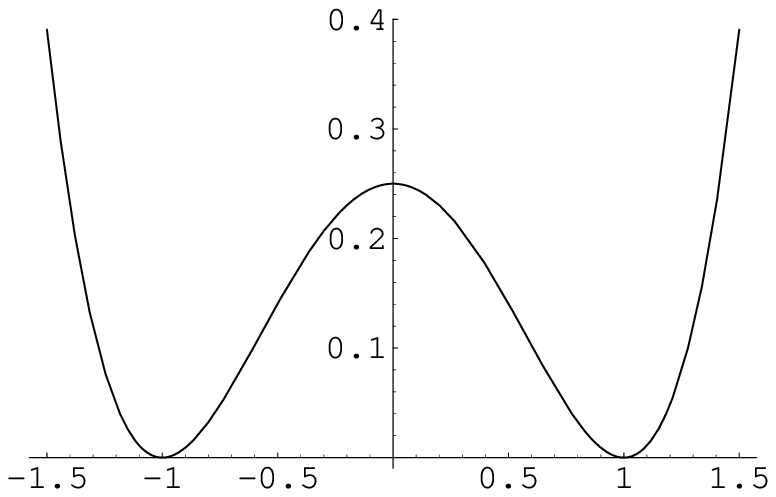}~a).~~~~~~~~~~~~
\includegraphics[width=43mm]{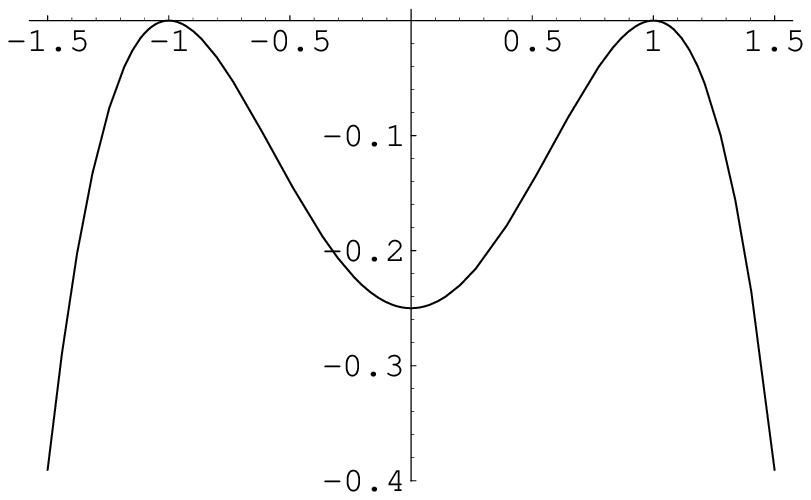}~b).
\caption{$W$ and $-W$ shaped potentials.}
\label{initial-potential}
\end{figure}

It is interesting to note, that initially we considered a model
with potential $V(\phi)=-\frac{1}{2}\phi^2+\frac{1}{4}\phi^4+\frac{1}{4}$,
see Fig. \ref{initial-potential}a).
Nonlocal term contributes to kinetic term, most importantly it
changes its sign \cite{IA_marion}, or equivalently changes the sign of potential
(kinetic term is positive in this case), see
Fig.\ref{initial-potential}. In this context it is interesting
to study already mentioned mechanical approximation which can
illustrate such an effect.

Solutions of (\ref{fr}) lead to interesting cosmological properties.
Let us consider them for physically interesting values of $\xi^2$
and see what kind of physical behavior they result in.
We would like to study dynamics of the following physical quantities.
The state parameter which is defined as usual
\begin{equation}
w=\frac{p}{\rho},
\end{equation}
or in the terms of Hubble parameter
\begin{equation}
w=-1-\frac{2}{3}\frac{\dot{H}}{H^2},
\end{equation}
will be presented on the figures below along with deceleration parameter
defined as
\begin{equation}
q(t)=-\frac{\ddot{a}a}{\dot{a}}.
\end{equation}

\subsection{Effective mechanical potential}
\label{mech-problem}

In order to better understand which behavior of the system we
might expect let us first study equation (\ref{eom-phi-H}) in
local approximations.

\begin{figure}[b!]
\centering
\includegraphics[width=53mm]{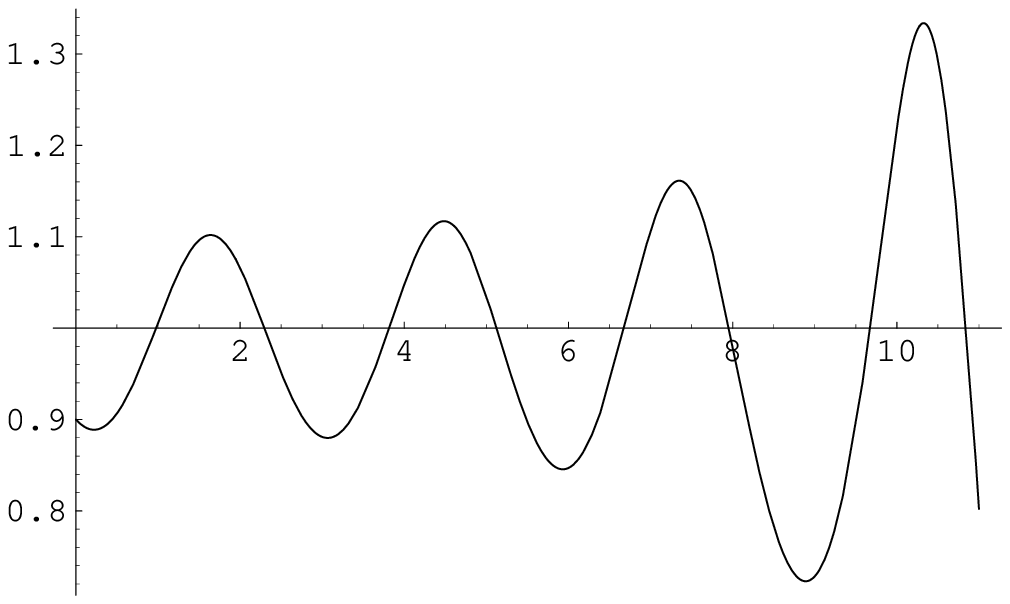}~a).~~~
\includegraphics[width=53mm]{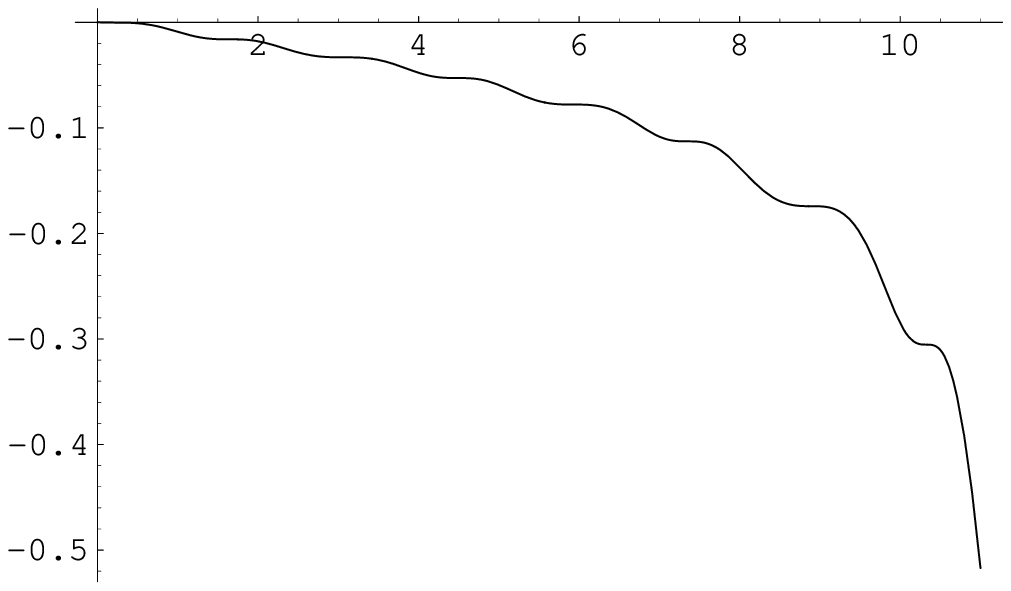}~b).~~~
\caption{Anharmonic oscillator with W-shape potential; a) periodic scalar field
trajectory located around $\Phi=1$; b) the solution for the Hubble function H(t),
$\xi^2=0.4$.}
\label{phih-approx1}
\end{figure}

If we ignore the nonlocal operator $e^{2k\Dh}$ in (\ref{eom-phi-H}) we get
\begin{equation}
\label{field-eq-app1}
(-\xi^2 \Dh+1)\Phi=\Phi^3.
\end{equation}
Let us reproduce action which leads to the equation above
\begin{equation}
\label{approx-act1}
S=\int d^4x\sqrt{-g}\left(\frac{m_p^2}{2}R+
 \frac{\xi^2}{2} \Phi\square_g\Phi+\frac{1}{2}\Phi^2-
\frac{1}{4}\Phi^4\right).
\end{equation}
This action describes a particle moving in the system with friction $H(t)$
\begin{subequations}
\begin{equation}
\xi^2\pd^2 \Phi=\Phi-\Phi^3-3H\xi^2\pd \Phi,
\end{equation}
\begin{equation}
\dot{H}=-\frac{\xi^2}{m_p^2}(\pd \Phi)^2.
\end{equation}
\end{subequations}
As we can see here Hubble function is explicitly negative and
thus we obtained a system which is unusual from mechanical
point of view -- it is an anharmonic oscillator with
\textit{negative} friction (see Fig. \ref{phih-approx1}).
Such a behavior looks similar with the behavior for
nonlocal systems studied in \cite{AJ-JHEP,Lump} in the
Minkowski background, when the trajectories of the scalar field
exceed the extremum of the potential, meanwhile the energy is
conserved.

\begin{figure}
\centering
\includegraphics[width=31mm]{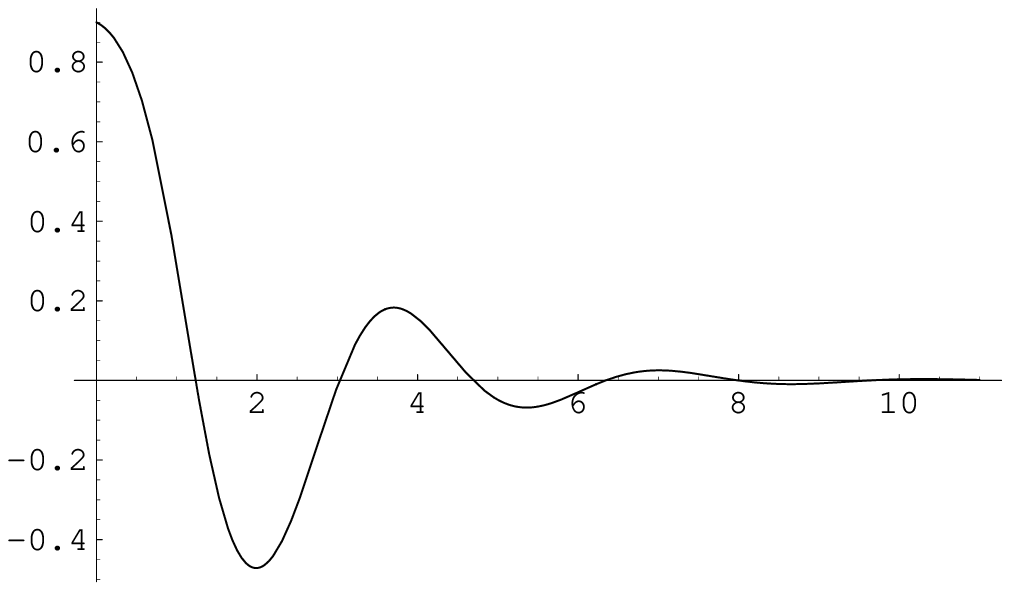}
\includegraphics[width=31mm]{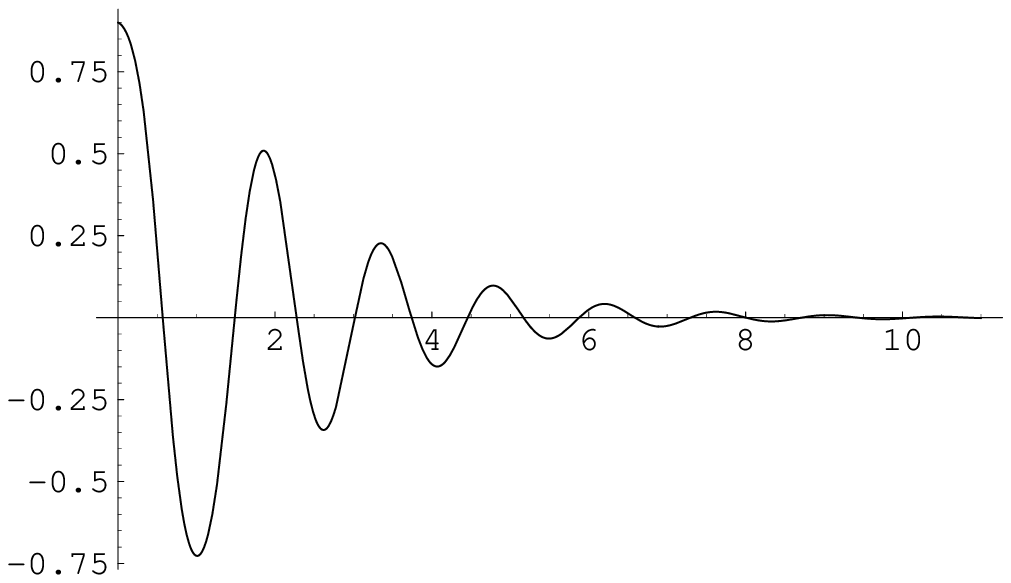}
\includegraphics[width=31mm]{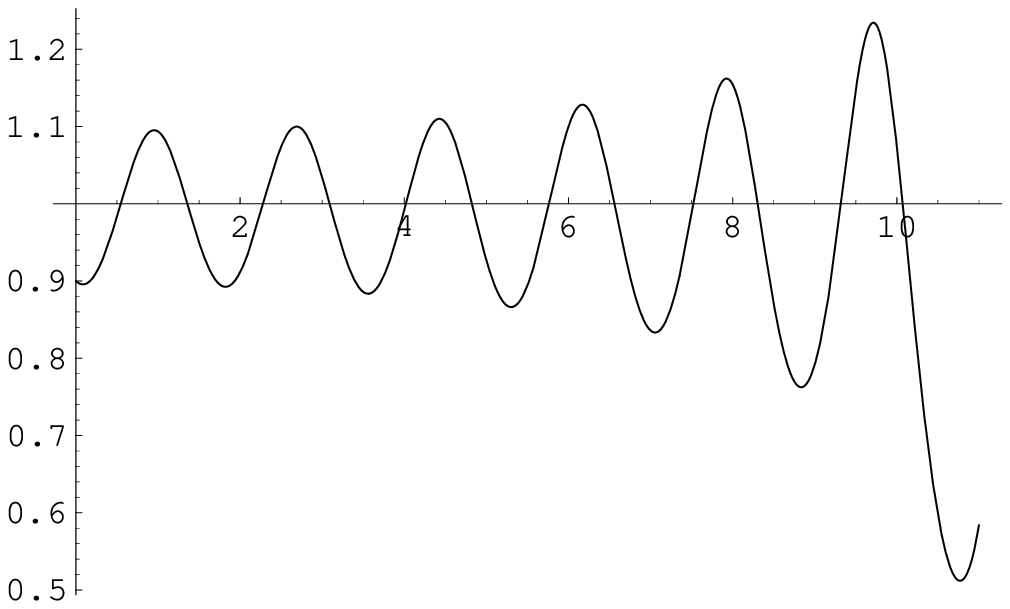}
\includegraphics[width=31mm]{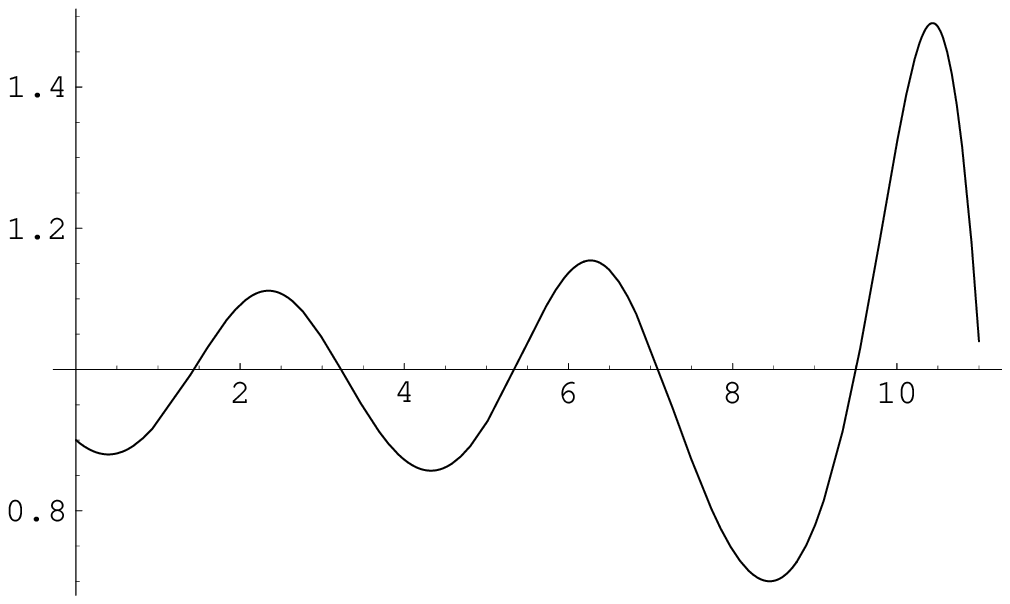}
\includegraphics[width=31mm]{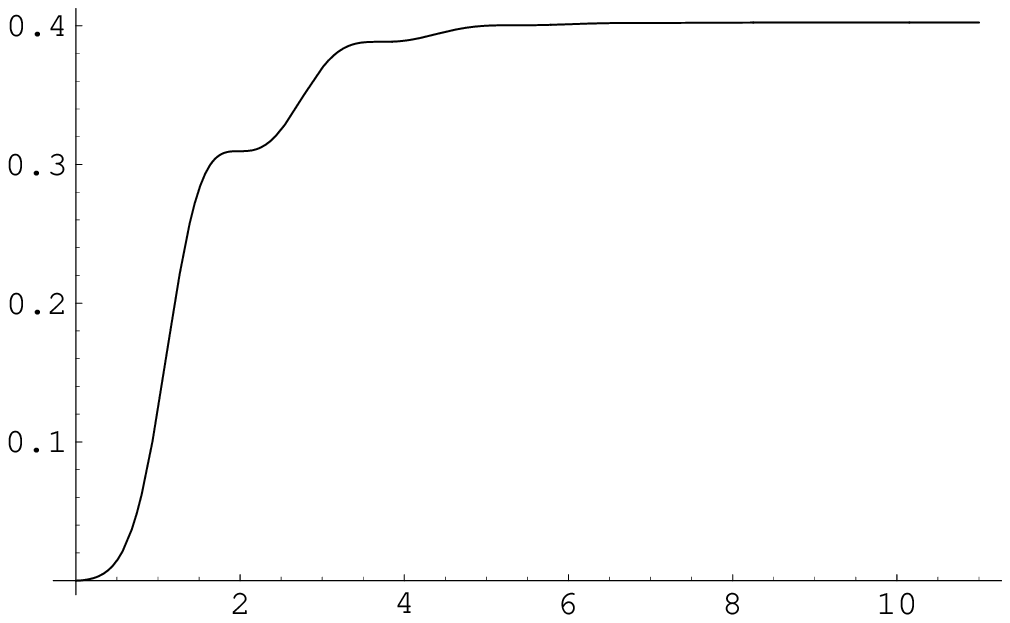}
\includegraphics[width=31mm]{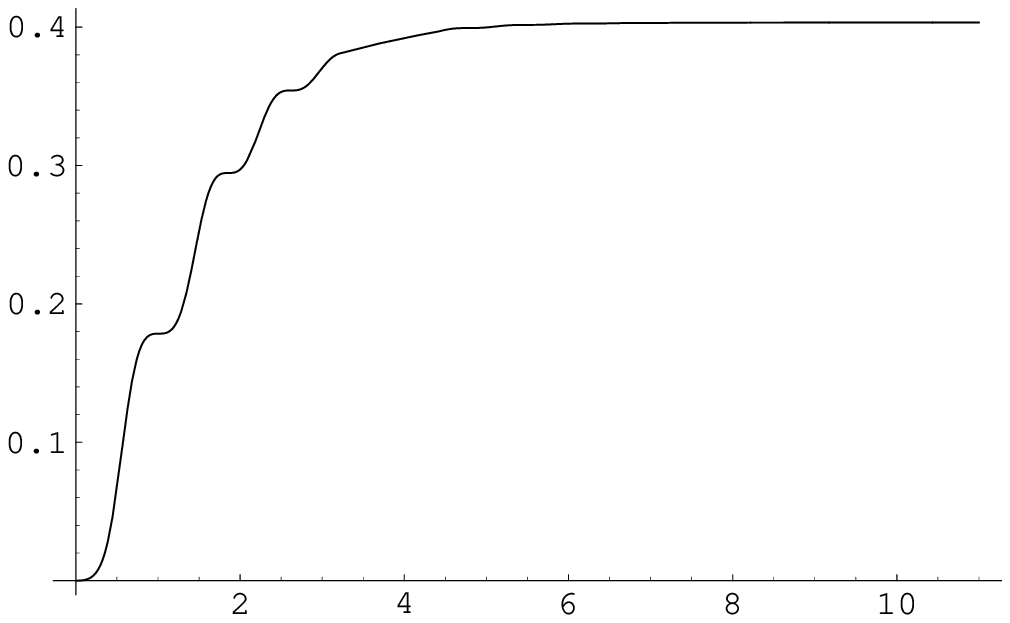}
\includegraphics[width=31mm]{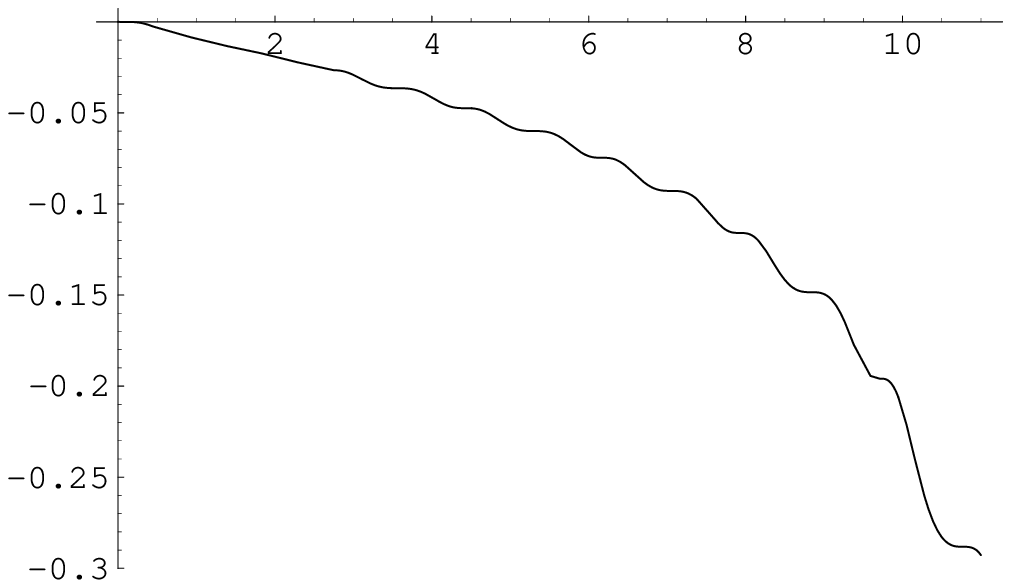}
\includegraphics[width=31mm]{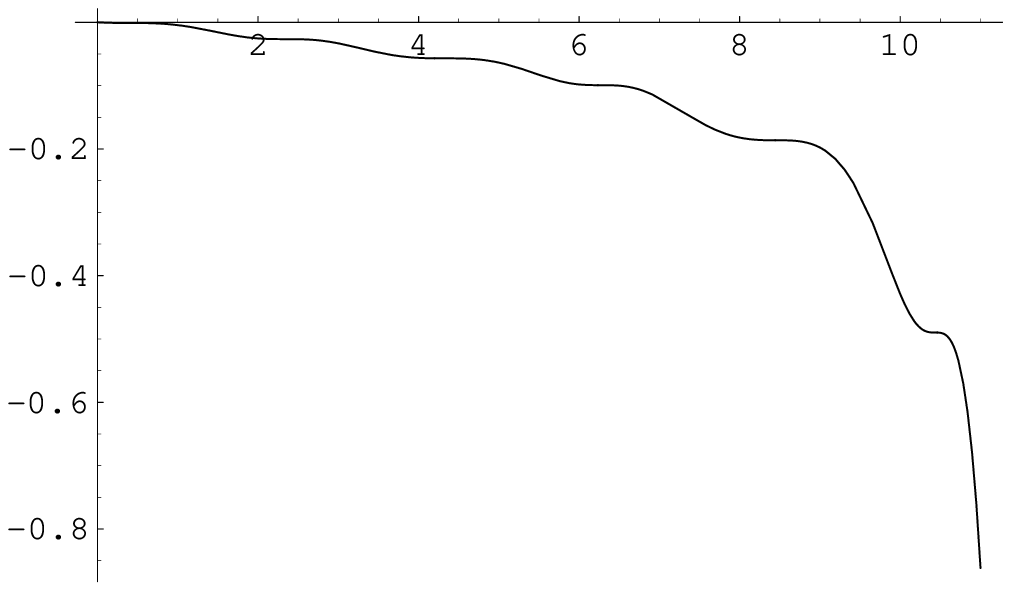}
\caption{Scalar field $\Phi(t)$ (upper row) and Hubble function
$H(t)$ (lower row) for the system (\ref{syst42}) for different
values for $\xi^2$ parameter $\xi^2=0, 0.2, 0.4, 1$ from left to
right correspondingly.}
\label{phih-approx2}
\end{figure}

Now let us consider mechanical approximation by accounting in
the equation (\ref{eom-phi-H}) only derivatives no higher that
second order, we get
\begin{equation}
\label{field-eq-app2}
((2k-\xi^2) \pd^2+1)\Phi=\Phi^3-3(2k-\xi^2)H(t)\pd \Phi.
\end{equation}
Equation (\ref{field-eq-app2}) could be obtain from the following action
\begin{equation}
\label{approx-act2}
S=\int d^4x\sqrt{-g}\left(\frac{m_p^2}{2}R+
 \frac{(\xi^2-2k)}{2} \Phi\square_g\Phi+\frac{1}{2}\Phi^2-
\frac{1}{4}\Phi^4\right),
\end{equation}
which leads to the following local Friedmann equations
\begin{subequations}
\begin{equation}
\pd^2 \Phi=-\frac{\Phi-\Phi^3}{(2k-\xi^2)}-3H\pd \Phi,
\end{equation}
\begin{equation}
\dot{H}=\frac{(2k-\xi^2)}{m_p^2}(\pd \Phi)^2.
\end{equation}
\label{syst42}
\end{subequations}
This system of equations describes particle moving in the
following potential
$$
V(\Phi)=\frac{2\Phi^2-\Phi^4}{4(2k-\xi^2)}.
$$
As we can see for $\xi^2<2k$ flip of the potential takes place.
In particular, for $\xi^2=0$ we get an equation describing
particle moving in the $-W$-shape potential (see
Fig.\ref{initial-potential}b). For $\xi^2 > 2k$ the potential
has $W$-shape (see Fig.\ref{initial-potential}a). This
simple remark leads to important physical consequences, we
might expect two different regimes of the solution.
Solutions for scalar field and Hubble function for different
values of $\xi^2$ are presented on Fig.\ref{phih-approx2}.
We see two regimes -- decreasing oscillations with positive
Hubble function and increasing oscillations with positive
Hubble function.

It is important to note that while mechanical approximation
helps us to illustrate or explain qualitatively some physical
consequences, it is a rather rough approximation and does not
necessarily capture some nonlocal effects. This situation is
already depicted by the fact that in full nonlocal model and
its mechanical approximation the change of regime of the
solution corresponding to $W$ and $-W$ shapes of the potential
happens for different values of $\xi^2$, they are of the same
order though.

\subsection{Rolling tachyon in the FRW background and the effective
cosmological constant $\Lambda^{\prime}$}

Let us study the value of effective cosmological
constant $\Lambda^{\prime}$. Formally $\Lambda^{\prime}$ enters
action as a correction of $D3$-brane tension $T$. For rolling
solution in Minkowski background it is known
\cite{Sen,AJK,IA_marion} that the value of $T$ is essential for
calculating stress tensor, but as it does not enter equations
of motion it is not essential for the existence of solution.
Moreover according to Sen's conjecture \cite{Sen,AJK,IA_marion}
$$
T_{\text{Minkowski}}=-V(\Phi=\pm 1)=\frac{1}{4}.
$$
Such value of $T$ in Minkowski background corresponds to zero
cosmological constant in the non-perturbative vacuum
$\Phi=\pm 1$.

In case of FRW metric we represent $D3$-brane tension as
$T=T_{minkowski}+\Lambda^{\prime}$ where $\Lambda^{\prime}$ is
determined uniquely for each field configuration (situation is
the same in local theories). Note that $\Lambda^{\prime}$ does
not enter equations (\ref{eqns}), in fact we determine its
value from (\ref{fr1}). Different values of $\Lambda^{\prime}$
for different $\xi$ are presented on table \ref{tab:L}. Note
that we worked with dimensionless variables. If we return to
initial notations and take into account physical constants in
action (\ref{full-action}) we can get realistic value of $\Lambda$
because generally speaking the string scale does not coincide with
the Plank mass \cite{IA_marion,BarnabyBiswasCline,AJV-JHEP}.

\begin{table}[h!]
\centering
\begin{tabular}{|c|c|c|c|c|c|c|}
\hline
$\xi^2$ & $\Lambda^{\prime}$\\
\hline
0       & 0.036938\\
\hline
0.2     &  0.010379\\
\hline
0.41    & 0.000022\\
\hline
0.42    & 0.\\
\hline
0.43    & 0.000023\\
\hline
0.6     & 0.007766 \\
\hline
0.96    & 0.077848\\
\hline
\end{tabular}
\caption{Values of $\Lambda^{\prime}$ for different $\xi^2$.}
\label{tab:L}
\end{table}

\subsection{$P$-adic cosmological model, $\xi^2=0$}

\begin{figure}[ht!]
\centering
\includegraphics[width=49mm]{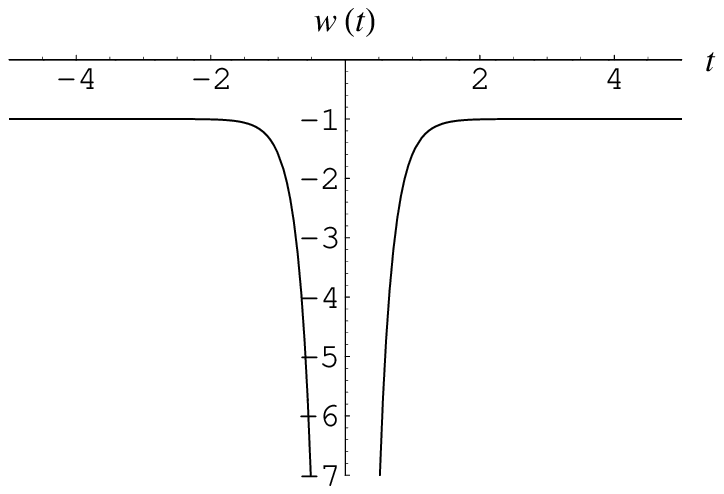}~~~
\includegraphics[width=49mm]{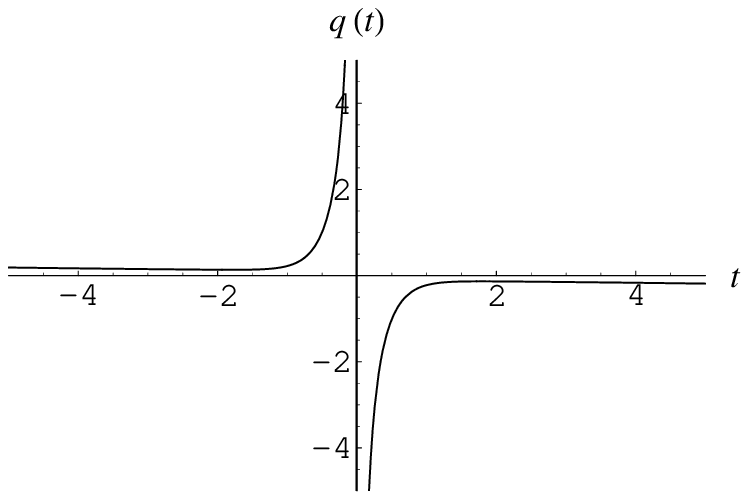}
\caption{The state and deceleration parameters for the case of
$p$-adic cosmological model ($\xi^2=0$).}
\label{xi0}
\end{figure}

Recently $p$-adic string models started to attract a lot of
attention being used as toy cosmological models. In
\cite{BarnabyBiswasCline} $p$-adic string model\cite{AJK}
was considered in the context of inflation and  approximate solutions of the
fully nonlocal $p$-adic string theory coupled to gravity were
constructed. In section \ref{ITforSS} we presented a new algorithm
for construction of precise solutions for these models. While
iterative procedure of section \ref{ITforSS} converges for any
odd $p$ (for rigorous mathematical proof for the simpler
Minkowski case see \cite{VladVol}) here we will analyze
cosmological properties of the solution for the case $p=3$.
This case is specifically interested since as mentioned earlier
it corresponds to zero mass of the tachyon field, $\xi^2=0$, in
level truncated fermionic string field theory (SFT).

We can see on Fig.\ref{H-iter} and Fig.\ref{xi0} that scalar
field and background solutions are monotonic functions which
tend to constants as $t\to\pm\infty$. Since these solutions do
not have turning points (oscillations) there is no crossing of
the cosmological state parameter barrier ($w=-1$), it is
negative and approaches to $-1$ from below (Fig.\ref{xi0}). For
these solutions we have nonsingular accelerating Universe with
a bounce. As an illustration we can note that behavior of the
scale factor obtained could be well approximated by
$a_{\text{approx}}(t)=e^{H_0t}$ for large $t>0$ and by
$a_{\text{approx}}(t)=e^{-H_0t}$ for large $t<0$, where
$H_0=\lim_{t\to\infty}H(t)$. Thus we get de Sitter type
solution for flat FRW background for $t\to+\infty$ and anti de
Sitter as $t\to-\infty$.
It might be useful to remark that bouncing cosmology is a subject of of recent
investigation which include the Ekpyrotic \cite{Turok}, Pre-Big-Bang \cite{Pre-Big-Bang-b-c}
and higher derivative
modification of Einstein gravity scenarios \cite{higher-der-mod-gr-b-c}.

Deceleration parameter $q$ is negative
and approaches $-1$ from below as $t\to\pm\infty$.

\subsection{SFT cosmological model, $\xi^2=\xip$}

\begin{figure}[ht]
\centering
\includegraphics[width=49mm]{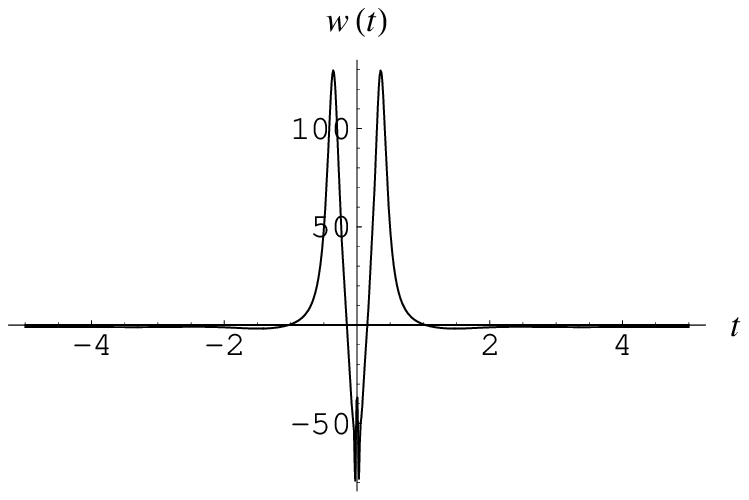}~~~
\includegraphics[width=49mm]{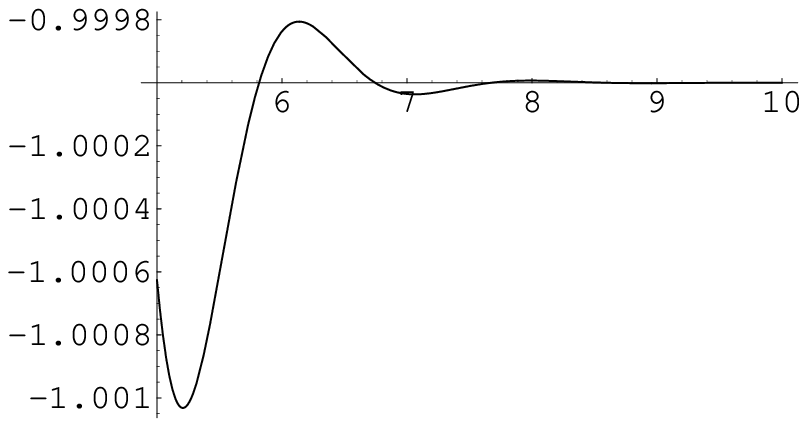}~~~
\includegraphics[width=49mm]{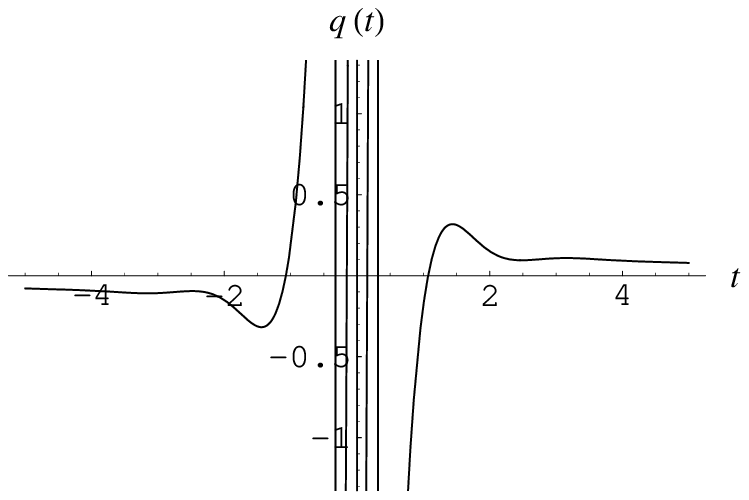}
\caption{The state parameter, its fine structure and
deceleration parameter for the case of SFT
cosmological model ($\xi^2=\xip$).}
\label{xi096}
\end{figure}

Today SFT is the strongest candidate for real physical theory
of nature. As we mentioned earlier level truncation in
non-sypersymmetric fermionic SFT leads to action (\ref{action})
with potential (\ref{potential}) with $\xi^2=\xip$
(\ref{xi2-phys}) \cite{ABKM,AJK,IA_marion}. The algorithm
presented in section \ref{ITforSS} allowed us to study this
model coupled to FRW backgroud without explicit approximations.

As shown on Fig.\ref{H-iter} and Fig.\ref{xi096} in this case
scalar field solution has turning points (decreasing
oscillations) and tends to stationary solutions $\pm 1$ as
$t\to\pm\infty$. The shape of Hubble function is essentially
different as compared to the case $\xi^2=0$, indeed it has a
clear maximum and tends to negative constant for $t>0$,
$\lim_{t\to\pm\infty}H(t)=\mp 0.15$. Such behavior is in
agreement with analysis presented in the previous section since
$\xip > \xio$.

The state parameter $w(t)$ tends to $-1$ as $t\to\pm\infty$ and
has exponentially decreasing oscillations around $w=-1$
barrier. Note that models which allow for crossing of $w=-1$
barrier are a subject of many recent investigations
\cite{Rubakov,Starobinsky,Mukhanov,Andrianov,Odintsov,Feng,Wei,AJV-JHEP}.
Here we observe this behavior in nonlocal model with only one
scalar field as in \cite{AJV-JHEP}. We also see that we have a
decelerating phase at late times, this property was not
predicted by approximate methods \cite{IA_marion}.

\section{Conclusions}

In this paper we have studied the dynamics of nonlocal
cosmological models driven by String Field Theory. These models
have an infinite number of derivatives and are characterized by
positive constants $k,~\xi$ and $\Lambda^{\prime}$ which
determine the shape of the solution. We have developed a new
method for solving nonlocal Friedmann equations. Such equations
contain infinite number of derivatives and form a new class of
equations in mathematical physics which recently started to be
discussed in literature
\cite{Yar-JPA,VladVol,LJ,Lump,BarnabyBiswasCline,Prokh,Vlad,CMN}.

To study cosmological properties we have investigated the
behavior of nonlocal models in the FRW background. We are
especially interested in two physical cases: p-adic
cosmological models and level-truncated SFT model. In $p$-adic
cosmological model we obtained nonsingular bouncing solution
with $w$ parameter approaching the $-1$ barrier from below. SFT
model has nonsingular solution for which $w$ parameter crosses
$-1$ barrier. We also discussed the possibility of obtaining
realistic cosmological constant from considered above nonlocal
cosmological models.

\section*{Acknowledgements}

The author would like to thank D. Mulryne and N. Turok for very
useful discussions. The author is also grateful to Ya. Volovich
for the assistance  with numerical calculations.
The author also would like to thank I.Ya.~ Aref'eva and S.V.~ Vernov for their
comments on an earlier version of this manuscript.
The author gratefully acknowledge the use of the UK National
Supercomputer, COSMOS, funded by PPARC, HEFCE and Silicon
Graphics. This work is supported by  the Center for Theoretical
Cosmology, in Cambridge and in part by RFBR grant 05-01-00758,
INTAS grant 03-51-6346  and Russian President's grant
NSh--672.2006.1.

\end{document}